\documentclass[aps,pre,preprint,groupedaddress,showpacs]{revtex4-1}
\usepackage{amssymb}

\usepackage[dvips]{graphicx}
\usepackage{textcomp}
\usepackage{amssymb}

\newcommand{\mc}{\multicolumn}

\begin{document}

\title{
\Large\bf Thermodynamic Casimir Effect in Films: 
          the Exchange Cluster Algorithm}

\author{Martin Hasenbusch}
\email[]{Martin.Hasenbusch@physik.hu-berlin.de}
\affiliation{
Institut f\"ur Physik, Humboldt-Universit\"at zu Berlin,
Newtonstr. 15, 12489 Berlin, Germany}

\date{\today}

\begin{abstract}
We study the thermodynamic Casimir force for films with  various types of 
boundary conditions and the bulk universality class of the three-dimensional
Ising model. To this end we perform Monte Carlo simulations of
the improved Blume-Capel model on the simple cubic lattice.  
In particular, we employ the exchange or geometric cluster
cluster algorithm [J.R. Heringa and H. W. J. Bl\"ote, Phys.\ Rev.\ E {\bf 57},  4976  (1998)].
In a previous work we demonstrated that this algorithm allows to compute
the  thermodynamic Casimir force for the plate-sphere geometry efficiently.
It turns out that also for the film geometry a substantial reduction of the
statistical error can achieved. Concerning physics, we focus on $(O,O)$ boundary
conditions, where $O$ denotes the ordinary surface transition. These are 
implemented by free boundary conditions on both sides of the film. 
Films with such boundary conditions undergo a phase transition in the universality 
class of the two-dimensional Ising model. We determine the inverse
transition temperature for a large range of thicknesses $L_0$ of the film and
study the scaling of this temperature with $L_0$. In the neighborhood
of the transition, the thermodynamic Casimir force is affected by finite size effects, 
where finite size refers to a finite transversal extension $L$ of the film.
We demonstrate that these finite size effects can be computed by using the universal
finite size scaling function of the free energy of the two-dimensional Ising model.
\end{abstract}

\pacs{05.50.+q, 05.70.Jk, 05.10.Ln, 68.15.+e}
\keywords{}
\maketitle

\section{Introduction}
In their seminal work, de Gennes and Fisher \cite{FiGe78} pointed out that the 
spatial restriction of thermal fluctuations should lead to an effective
force. Due to its analogy with the Casimir effect \cite{Casimir48}, where 
the spatial restriction of quantum fluctuations leads to a force, it is called
thermal, thermodynamic or critical Casimir effect. Here ``critical'' refers to the 
fact that thermal fluctuations become large in the neighbourhood of a critical
point. At a second order phase transition, in the thermodynamic limit of the bulk 
system, the correlation length, which characterizes the spatial extent of
these fluctuations, behaves as
\begin{equation}
 \xi \simeq \xi_{0,\pm} t^{-\nu} \;,
\end{equation}
where $\xi_{0,\pm}$ are the amplitudes of the correlation length in the high and the 
low temperature phase, respectively, and $\nu$ is the critical exponent of the 
correlation length. 
The reduced temperature is given by $t = (T-T_c)/T_c$, where $T_c$ is the critical temperature. Note that
in the following we shall use for simplicity $t = \beta_c-\beta$, where $\beta = 1/k_B T$.
For reviews on critical phenomena see for example \cite{WiKo,Fisher74,Fisher98,PeVi02}.

Owing to their simplicity, often films are studied. For films the thermodynamic Casimir
force per area is given by
\begin{equation}
\label{defineFF}
 F_{Casimir} = -   \frac{ \partial \tilde f_{ex} }{ \partial L_0} \;\;,
\end{equation}
where $\tilde f_{ex} = \tilde f_{film} - L_0 \tilde f_{bulk}$
is the excess free energy per area of the film of thickness $L_0$, 
where $\tilde f_{film}$ is the free energy per area of the film
and $\tilde f_{bulk}$ is the free energy density of the bulk system.
The thermodynamic Casimir force per area follows the finite size scaling law
\begin{equation}
\label{FCffs}
 F_{Casimir} \simeq k_B T L_0^{-3} \; \theta(t [L_0/\xi_{0,+}]^{1/\nu}) \;\;,
\end{equation}
see for example ref. \cite{Krech}. The function $\theta$ is expected to be universal, 
which means that it should only depend on the universality classes of the transitions
of the bulk system and the surfaces. For reviews on surface critical phenomena
see \cite{BinderS,Diehl86,Diehl97}.

The thermodynamic Casimir effect has been demonstated in experiments 
on films of  $^4$He and $^3$He-$^4$He mixtures near the $\lambda$-transition or
the tri-critical point
of the bulk system \cite{GaCh99,GaCh02,UeBaMiCaRo03,GaScGaCh06}. The force obtained for 
different thicknesses is described quite well by a unique scaling function $\theta(x)$.
Also experiments with liquid binary mixtures near the mixing-demixing
transition were performed, where either
films \cite{FuYaPe05,RaBoJa07} or the sphere-plate geometry 
\cite{Nature,So08,GaMaHeNeHeBe09,NeHeBe09,Tr11,Nellen,Zvyagolskaya}
were  studied.
In other experiments, the thermodynamic Casimir force is the driving force
for colloidal aggregation \cite{Bonnetal09,ZvArBe11}.

It is a theoretical challenge to compute the universal scaling function $\theta(x)$
for different bulk universality classes and types of boundary conditions to compare
with experimental data. Still the mean-field approximation is used as tool that can 
be employed relatively easily for more complicated geometrical setups. For recent 
work see for example \cite{PaTrDi13,MaHaDi14}. Obviously, no accurate 
results can be expected this way. Unfortunately field theoretic methods do not
allow to compute $\theta(x)$ for all types of boundary conditions of interest
or do not allow to compute $\theta(x)$ in the full range of the scaling variable $x$
\cite{KrDi1,KrDi2,DiGrSh06,GrDi08,DiGr09,ScDi08,DiSc11,Dohm09,Dohm11,Dohm13,Dohm14}. 
For a discussion of this point see 
for example the introduction of \cite{DiGrHaHuRuSc14}.  Exact results can be obtained
in the large $N$ limit for periodic and free boundary conditions 
\cite{DiGrHaHuRuSc14,Da96,Da98,ChDa04,DaDiGr06,DaGr09,DiGrHaHuRuSc12,DaBeRu14}. 
Also for the two-dimensional Ising model with various boundary conditions exact 
results were obtained \cite{EvSt94,NoNa,AbMa10,RuZaShAb10,WuIzGu12,AbMa13}. 
In the case
of the three-dimensional Ising universality class and strongly symmetry breaking boundary conditions,
quite accurate results had been obtained by using  the extended de Gennes-Fisher  
local-functional method  \cite{BoUp98,BoUp08,UpBo13}. O($n$)-symmetric systems with periodic
boundary conditions had been studied using a functional renormalization group approach \cite{JaNa13}.

In the last few years there has been considerable progress in the  
study of the thermodynamic Casimir force by using Monte Carlo simulations of 
lattice spin models. At least in princible, the finite size scaling function
can be determined with a controlable statistical and systematical error.
In particular, in refs. \cite{Hucht,VaGaMaDi07,VaGaMaDi08,Ha09,Ha10}
the three-dimensional XY bulk universality class and a vanishing field at 
the boundary have been studied, which is relevant for the experiments on $^4$He.
A quite satisfactory agreement between the experimental results and the theory was
found. In refs.
\cite{DaKr04,VaGaMaDi07,VaGaMaDi08,MHstrong,PaToDi10,mycrossing,VaMaDi11,HuGrSc11,MHcorrections,VaDi13,PaTrDi13,CaJaHo14,Va14,PaTrDi14}
the Ising bulk universality class and various types of boundary conditions were studied.
Note that a continuous mixing-demixing transition of binary mixtures 
belongs to the Ising bulk universality class.
Notwithstanding this nice progress, further algorithm improvements are certainly welcome to 
study problems with a large parameter space like structured surfaces \cite{PaTrDi13,PaTrDi14},
disorder at the surface, the crossover from the special to the ordinary surface 
universality class \cite{ScDi08}, the presence of an external bulk field \cite{VaDi13,CaJaHo14}, or more complicated geometrical setups \cite{MaHaDi14}.

In ref. \cite{mysphere} we determined the  thermodynamic Casimir force
for the plate-sphere geometry. We studied the three-dimensional Ising universality
class and strongly symmetry breaking boundary conditions. A preliminary 
study showed that with a conventional approach and a reasonable amount of 
CPU-time it is impossible to get meaningful results for this problem.
Employing the exchange cluster algorithm, it is possible to define a variance
reduced estimator for the difference of the internal energy. This allowed 
us to obtain the scaling functions of the thermodynamic Casimir force with 
high accuracy.
The exchange cluster algorithm is a variant of the geometric cluster algorithm 
of \cite{HeBl98}. In the geometric cluster algorithm the sites of a single lattice 
are organized in pairs. This is achieved for example by a reflection at a plane
of the lattice. The elementary operation of the update is the exchange of the 
spin value within such pairs of sites. Instead, we consider two independent 
systems. 
We consider 
pairs of sites, where one is in one lattice, while the other site belongs to 
the other lattice.  

In the present work we apply the exchange cluster algorithm to the
film geometry. 
The relative simplicity of the film geometry allows us to study 
the properties of the exchange cluster algorithm and its associated 
improved estimators more systematically.  
In the present work we first study strongly symmetry breaking boundary 
conditions, $(+,+)$ and $(+,-)$, then  $(+,O)$ and finally $(O,O)$ 
boundary conditions. Here the sign
indicates the value of the spins at the boundary and $O$  the 
ordinary surface transition. These problems have been studied before, 
and the scaling functions of the thermodynamic Casimir force are 
known fairly well. Here we are mainly aiming at a better understanding
of the exchange cluster algorithm before attacking more complicated
problems.  It turns out that,
depending on the type of the surfaces of the film, large
reductions of the variance can be achieved.

In the case of 
$(O,O)$ boundary conditions, the problem arises that the film 
undergoes a second order phase transition in the universality class 
of the two-dimensional Ising model. This leads to sizeable finite size
effects, where the  finite extension in the transversal directions is
meant. To understand these finite size effects and the interplay of 
the transition with 
the thermodynamic Casimir force on a quantitative level, we first 
accurately determined the critical temperature for a large range of 
thicknesses $L_0$ by using the method discussed in  \cite{CaHa96}.
We match the reduced temperature of the two-dimensional Ising model and the films.
We analyze how the temperature of the effectively two-dimensional transition 
approaches the bulk transition temperature as the thickness of the 
film increases.

Based on these results, we demonstrate that finite size effects
of the thermodymanic Casimir force due to the finite extension of the
lattice in the transversal directions are governed by the universal
finite size scaling function of the free energy density
that is obtained by analyzing the two-dimensional Ising model.

The paper is organised as follows. In section \ref{themodel} we define the model and discuss the 
boundary conditions that we study in this work.
In section \ref{algorithm} we discuss the exchange cluster algorithm and the 
variance reduced estimator for differences of the internal energy and 
other quantities. At the example of $(+,-)$ boundary conditions at the critical point
of the bulk system, we carefully study how the performance of the algorithm depends
on its parameters.
In sections \ref{strongsection} and \ref{mixedresults} we present our numerical results for 
strongly symmetry breaking and $(O,+)$ boundary conditions, respectively.
In section \ref{finite2D} we determine the finite size scaling function of the free energy
density of the two-dimensional Ising model.
In section \ref{filmtc} we study the phase transition of films with $(O,O)$ boundary conditions
for a large range of thicknesses $L_0$. Then in section \ref{casimiroo} we determine the scaling function of
the thermodynamic Casimir force for films with $(O,O)$ boundary conditions.
Finally we summarize our results and give an outlook.
\section{The model}
\label{themodel}
As in previous work, we study the Blume-Capel model on the simple cubic lattice.
The bulk system, in absence of an external field, is defined by the reduced Hamiltonian
\begin{equation}
\label{Isingaction}
H = -\beta \sum_{<xy>}  s_x s_y
  + D \sum_x s_x^2  \;\; ,
\end{equation}
where the spin might assume the values $s_x \in \{-1, 0, 1 \}$.
$x=(x_0,x_1,x_2)$
denotes a site on the simple cubic lattice, where $x_i \in \{1,2,...,L_i\}$
and $<xy>$ denotes a pair of nearest neighbors on the lattice.
The inverse temperature is denoted by $\beta=1/k_B T$. The partition function
is given by $Z = \sum_{\{s\}} \exp(- H)$, where the sum runs over all spin
configurations. The parameter $D$ controls the
density of vacancies $s_x=0$. In the limit $D \rightarrow - \infty$
vacancies are completely suppressed and hence the spin-1/2 Ising
model is recovered.

In  $d\ge 2$  dimensions the model undergoes a continuous phase transition
for $-\infty \le  D   < D_{tri} $ at a $\beta_c$ that depends on $D$, while
for $D > D_{tri}$ the model undergoes a first order phase transition,
where $D_{tri}=2.0313(4)$, see ref. \cite{DeBl04}.

Numerically, using Monte Carlo simulations it has been shown that there
is a point $(D^*,\beta_c(D^*))$
on the line of second order phase transitions, where the amplitude
of leading corrections to scaling vanishes. In \cite{MHcritical} we
simulated the model at $D=0.655$ close to $\beta_c$ on lattices of a
linear size up to $L=360$. We obtained $\beta_c(0.655)=0.387721735(25)$ 
and $D^*=0.656(20)$.
The amplitude of leading corrections to scaling at $D=0.655$ is at
least by a factor of $30$ smaller than for the spin-1/2 Ising model.
Following eq.~(12) of ref. \cite{MHcorrections}, the amplitude 
of the second moment correlation length in the high temperature phase at $D=0.655$
is
\begin{eqnarray}
\label{xi0}
\xi_{2nd,0,+} &=&  0.2283(1) - 1.8 \times (\nu-0.63002)
                        + 275 \times (\beta_c - 0.387721735) \;\; \nonumber \\
&&  \mbox{using} \;\; t = \beta_c - \beta \;\; 
 \mbox{as definition of the reduced temperature} .
\end{eqnarray}
In the high temperature phase there is little difference between
$\xi_{2nd}$ and the exponential correlation length $\xi_{exp}$ which
is defined by the asymptotic decay of the two-point correlation function.
Following  \cite{pisaseries}:
\begin{equation}
\lim_{t\searrow 0} \frac{\xi_{exp}}{\xi_{2nd}} = 1.000200(3)
\;\;
\end{equation}
for the thermodynamic limit of the three-dimensional system.
Note that in the following $\xi_{0}$ always refers to $\xi_{2nd,0,+}$.

\subsection{Film geometry and boundary conditions}
\label{filmgeometry}
In the present work we study the thermodynamic Casimir effect
for systems with film
geometry. In the ideal case this means that the system has a finite
thickness $L_0$, while in the other two directions the
limit $L_1, L_2 \rightarrow \infty$ is taken. In our  Monte Carlo
simulations we shall study lattices with $L_0 \ll L_1, L_2$ and
periodic boundary conditions in the $1$ and $2$ directions.  Throughout
we simulate lattices with $L_1=L_2=L$.

The types of boundary conditions discussed here can be characterized
by the reduced Hamiltonian
\begin{equation}
\label{Isingaction2}
H = - \beta \sum_{<xy>}  s_x s_y + D \sum_x s_x^2   
\;-\; h_1 \sum_{x,x_0=1} s_x
\;-\; h_2 \sum_{x,x_0=L_0} s_x  \;\;,
\end{equation}
where $h_1, h_2 \ne 0$ break the symmetry at the surfaces.
In our convention $<xy>$ runs over
all pairs of nearest neighbor sites. Note
that here the sites $(1,x_1,x_2)$ and $(L_0,x_1,x_2)$ are not nearest
neighbors as it would be the case for periodic boundary conditions.
In general there is ambiguity, where exactly 
the boundaries are located and how the thickness of the film is precisely
defined. Here we follow the convention that $L_0$ gives the number
of layers with fluctuating spins.

First we study strongly symmetry breaking boundary 
conditions that are given by $|h_1|$, $|h_2| \rightarrow \infty$.
There are, up to symmetry transformations, two choices. 
Either $h_1$ and $h_2$ have the same or a different sign,
which we shall denote by $(+,+)$ and $(+,-)$, respectively.
Taking the limit $|h_1|$, $|h_2| \rightarrow \infty$ fixes 
the spins at the surface to the sign of the surface field.

In order to keep $L_0$ layers of fluctuating spins, which is done 
to be consistent with our previous work \cite{MHstrong,MHcorrections}, we actually 
put the surface fields $|h_1|=|h_2| \rightarrow \infty$ at $x_0=0$ and $x_0=L_0+1$.
Note that this is equivalent to $|h_1|=|h_2|=\beta$ at $x_0=1$ and $x_0=L_0$.
In a semi-infinite system, following the classification of refs.
\cite{BinderS,Diehl86,Diehl97}, this choice of boundary conditions corresponds to 
the normal or extraordinary surface universality class.

Next we simulated the case $h_1=0$ at $x_0=1$ and $h_2 \rightarrow \infty$ at $x_0=L_0+1$. 
In a semi-infinite system, a vanishing external surface field corresponds 
to the ordinary surface universality class. Hence, we denote this combination 
of boundary conditions by $(O,+)$.  
Finally we simulated systems with $h_1=0$ and $h_2=0$ at $x_0=1$ and $x_0=L_0$. This set of 
boundary conditions is denoted by $(O,O)$.  
In our program code we have implemented these boundary conditions
by spin variables that reside at $x_0=0$ and $x_0=L_0+1$ that are fixed to 
either $-1$, $0$, or $1$, depending on the type of the boundary condition.

In the case of $(O,+)$ and $(O,O)$
boundary conditions, we studied small $h_1$ and $h_2$ by computing the coefficients
of the Taylor-expansion of the quantities of interest up to second order around 
vanishing surface fields.

Given that leading bulk corrections are eliminated, the leading remaining corrections
are due to the surfaces. There are theoretical arguments that these can be expressed 
by an effective thickness $L_{0,eff}=L_0 + L_s$ of the film \cite{CaFi76}. 
The value of $L_s$ depends
on the precise definition of the thickness $L_0$. Concerning the physics, it depends on 
the model that is considered and the type of boundary conditions that are imposed.
However it should be independent of the scaling variable $x$ and the physical quantity 
that is considered. It can be decomposed as $L_s=l_{ex,1} + l_{ex,2}$, where 
$l_{ex,i}$ are extrapolation lengths that depend on the type of boundary conditions
at the boundary $i$ and not on the boundary conditions at the other boundary.
For a discussion see for example section IV of \cite{MHstrong} or section III of \cite{PaTrDi13}.
In ref. \cite{DiGrHaHuRuSc14} the concept of an effective thickness has been 
verified with high numerical precision for the large $N$ limit of the three-dimensional
$O(N)$-symmetric $\phi^4$ model with free boundary conditions.  In the following we
shall use the numerical values $L_s=1.91(5)$, ref.  \cite{MHcorrections}, for strongly
symmetry breaking boundary conditions,  $L_s=1.43(2)$ for $(O,+)$ boundary conditions \cite{mycrossing}.
In the case of $(O,O)$ we take $L_s=2 l_{ex,O}$ where $l_{ex,O}=0.48(1)$, see eq.~(63) of
\cite{mycrossing}. The estimates of $L_s$ were obtained by analyzing the finite size scaling 
behavior of various quantities directly at the critical point.  Analyzing  the numerical
results for the thermodynamic Casimir force below, we shall use these values as input.

\section{Computing the thermodynamic Casimir force}
The reduced excess free energy per area of the film is defined by
\begin{equation}
f_{ex} = -\frac{1}{L_1 L_2} \ln Z  - L_0 f_{bulk}
\end{equation}
where $f_{bulk}$ is the reduced bulk free energy density and $Z=\sum_{\{s\}} \exp(-H(\{s\}))$ is the 
partition function of the film.
We compute the thermodynamic Casimir force by using eq.~(\ref{defineFF}). On the lattice,
the partial derivative of the reduced excess free energy per area with respect to the thickness of the film
is approximated by
\begin{equation}
\label{discr}
\frac{ \partial  f_{ex} }{ \partial L_0} \simeq \Delta f_{ex} = \frac{f_{ex}(L_0+d/2) - f_{ex}(L_0-d/2)}{d}
\end{equation}
where $d$ is a small positive integer. Except for a few preliminary algorithmic 
studies, we shall use the minimal value $d=1$.  Following Hucht \cite{Hucht},
we compute the difference of free energies as integral over the inverse temperature
of the difference of the corresponding internal energies
\begin{equation}
 \Delta f_{ex}(\beta)  = \Delta f_{ex}(\beta_0) 
- \int_{\beta_0}^{\beta} \mbox{d} \tilde \beta \Delta E_{ex}(\tilde \beta)
\end{equation}
where $\Delta E_{ex} = \langle \Delta E \rangle - E_{bulk}$ and 
\begin{equation}
\Delta E = \frac{E(L_0+d/2) - E(L_0-d/2)}{d}  
\end{equation}
where in our convention the energy per area is given by
\begin{equation}
 E = \frac{1}{L_1 L_2} \sum_{<xy>} s_x s_y 
\end{equation}
and $E_{bulk}$ is the bulk energy density. 
The integration is done numerically, using
the trapezoidal rule: 
\begin{equation}
\label{integration}
-\Delta f_{ex}(\beta_n) \approx
-\Delta f_{ex}(\beta_0) +
\sum_{i=0}^{n-1} \frac{1}{2} (\beta_{i+1}-\beta_i)
   \left[\Delta E_{ex}(\beta_{i+1}) + \Delta E_{ex}(\beta_{i}) \right]
\end{equation}
where $\beta_{i}$ are the values of $\beta$ we simulated at. They
are ordered such that $\beta_{i+1} > \beta_i$ for all $i$. Typically
$O(100)$ nodes $\beta_i$ are needed to compute the thermodynamic Casimir force in the whole
range of temperatures that is of interest to us.
Obviously, $\Delta f_{ex}(\beta_0)$ should be known with good accuracy. Usually one chooses
$\beta_0$ such that $\xi_{bulk}(\beta_0) \ll L_0$ and hence  $\Delta f_{ex}(\beta_0) \approx 0$. In the 
case of strongly symmetry breaking boundary conditions, we shall use a different choice of $\beta_0$ that 
is discussed in \cite{MHstrong,MHcorrections}. 

One important aspect of the present work is to demonstrate that the exchange cluster 
algorithm allows to compute $\langle \Delta E \rangle$ by using a variance reduced estimator.
The reduction of the variance depends on the type of the boundary conditions and 
the parameters $L_0$, $d$ and $\beta$ as we shall see below.
The variance of $\Delta E$, computed in the standard way, is 
\begin{equation}
 \mbox{var}(\Delta E)=\frac{\mbox{var}(E(L_0+d/2))+\mbox{var}(E(L_0-d/2))}{d^2}
               \approx  \frac{2 \mbox{var}(E(L_0))}{d^2}  \;\;.
\end{equation}
At the critical point, taking $L_1$ and $L_2$ proportional to $L_0$, the variance 
of the energy per area behaves as
\begin{equation}
\mbox{var}(E(L_0)) \propto C(L_0) L_0^{-1}
\propto L_0^{-1 + \alpha/\nu} = L_0^{-4 + 2/\nu}
\end{equation}
where $C(L_0)$ is the specific heat of the finite system. On the 
other hand, the quantity we are interested in scales as
\begin{equation}
 \Delta E_{ex} \propto  L_0^{-3 + 1/\nu}
\end{equation}
at the critical point. Hence the ratio
\begin{equation}
\label{problemvar}
\frac{\mbox{var}(\Delta E)}{(\Delta E_{ex})^2} \propto \frac{L_0^2}{d^2}
\end{equation}
which is, for a given number of statistically independent measurements,
proportional to the square of the statistical error, increases with
increasing thickness $L_0$. In order to keep the statistical error
small,  we used in ref.  \cite{MHcorrections} $d=2$ and $4$
for $L_0=33$ and $L_0=66$, respectively. This in turn makes it more
difficult to control the discretization error of eq.~(\ref{discr}).
As we shall see below, the exchange cluster improved estimator of
$\langle \Delta E \rangle$ eliminates this problem and for strongly symmetry breaking boundary conditions,
we get statistically accurate results for $L_0=64.5$ and $d=1$.
Note that, with comparable numerical effort, $E_{bulk}$ can be computed more accurately than 
$\langle \Delta E \rangle$, even when using the exchange cluster improved estimator. 
Here we shall mainly use numerical results for $E_{bulk}$ obtained in previous work
\cite{MHstrong,MHcorrections}. For a discussion, see section VII of \cite{MHstrong}. 
Note that one could also simulate the geometry discussed in ref. \cite{HoHu14} by using the
exchange cluster algorithm exactly in the same fashion as we simulated the sphere-plate geometry
in ref. \cite{mysphere}. The layer of fixed spins, called ``wall'' by the authors, which separates
two sub-systems, would take over the  role of the sphere. This way the simulation allows to measure
$\Delta E_{ex}$ directly.  Effectively $E_{bulk}$ is provided by the larger of the two sub-systems.
We performed a preliminary study that demonstrated that this indeed works. However we did not
follow this line, since, as discussed above, accurate results for $E_{bulk}$ are already available 
from  simulations of systems with periodic boundary conditions in all directions.

\section{The exchange cluster algorithm}
\label{algorithm}
With the exchange cluster algorithm, we simulate two systems that are defined on identical lattices.
Let us denote the sites of this pair of lattices by $s_{x,i}$, where $x$ labels a site in a given 
lattice and $i \in \{1,2\}$ denotes the lattice.
The sites of these two lattices are mapped by $T(x)$ one to one on each other such that the neighborhood
relation of the sites is preserved. In the simplest case, 
$T(x)$ is the identity. Here we shall use random translations along the transversal directions
of the film. One also could employ reflections.

The basic operation of the exchange cluster algorithm is to exchange the values of the spins 
between corresponding sites.
This operation can be described by an auxiliary variable $\sigma_x \in \{-1,1\}$:
\begin{eqnarray}
 \tilde s_{x,1} &=& \frac{1 +\sigma_x}{2} s_{x,1} + \frac{1 -\sigma_x}{2} s_{x,2}  \;, \\
 \tilde s_{x,2} &=& \frac{1 +\sigma_x}{2} s_{x,2} + \frac{1 -\sigma_x}{2} s_{x,1}  \;.
\end{eqnarray}
In order to keep the notation simple, we assume $T(x)=x$.
For $\sigma_x=-1$ the exchange is performed, while for $\sigma_x=1$ the old values are kept.
The contribution of a pair $<xy>$ of nearest neighbors to the reduced Hamiltonian is given by
\begin{eqnarray}
H_{<xy>} &=&  - \beta (\tilde s_{x,1} \tilde s_{y,1} + \tilde s_{x,2} \tilde s_{y,2})  \nonumber \\
 &=& - \frac{\beta}{2} \left( s_{x,1} - s_{x,2} \right)  \left( s_{y,1} - s_{y,2} \right) \sigma_x \sigma_y 
     - \frac{\beta}{2} \left( s_{x,1} + s_{x,2} \right)  \left( s_{y,1} + s_{y,2} \right) \;\;. \label{Hem}
\end{eqnarray}
Note that  terms linear in $\sigma$ cancel.  
The exchange of spins is performed by using a cluster update. The construction of the clusters
is characterized by the probability to delete the link between the nearest neighbors $x$ and $y$
\cite{HeBl98}
\begin{equation}
\label{pdsimple}
 p_d = \mbox{min} [1, \exp(-2 \beta_{embed} )]  \;,
\end{equation}
where 
\begin{equation}
\beta_{embed} = \frac{\beta}{2} ( s_{x,1} - s_{x,2} )  ( s_{y,1} - s_{y,2} ) \;,
\end{equation}
which is the prefactor of $\sigma_x \sigma_y$ in eq.~(\ref{Hem}). This is sufficient for the 
problems studied in this work. Let us briefly sketch how the exchange cluster algorithm can be applied 
to a more general class of problems.
For an enhanced coupling at the boundary, as it is required for the study of the special surface
universality class, eq.~(\ref{Hem}) has to be generalized to
\begin{equation}
 H_{<xy>} =  - \beta_{<xy,1>} \tilde s_{x,1} \tilde s_{y,1} - \beta_{<xy,2>} \tilde s_{x,2} \tilde s_{y,2}
\;.
\label{Hem2}
\end{equation}
This leads to the embedded coupling
\begin{equation}
\beta_{<xy>,embed} = \frac{\beta_{<xy,1>} + \beta_{<xy,2>}}{4} ( s_{x,1} - s_{x,2} )  ( s_{y,1} - s_{y,2} )
\end{equation}
and in addition to an external field that acts on $\sigma$:
\begin{equation}
 h_{x,<xy>,embed} = \frac{\beta_{<xy,1>} - \beta_{<xy,2>}}{4}  (s_{x,1} - s_{x,2}) ( s_{y,1} + s_{y,2} )
\;,
\end{equation}
where the indices of $h$ indicate that it is the contribution to the field at the site $x$ stemming 
from the pair $<xy>$  of sites. In case there is also an external field in the original problem 
we get the contribution 
\begin{equation}
h_{x,x,embed} = \frac{h_{x,1} - h_{x,2}}{2} (s_{x,1} - s_{x,2}) \;\;.
\end{equation}
In total
\begin{equation}
 h_{x,embed} = h_{x,x,embed} + \sum_{y.nn.x} h_{x,<xy>,embed} \;\;.
\end{equation}
This generalized problem can be simulated for example by constructing the clusters only based
on the pair interaction and then taking into account the external field in the probability to flip
the cluster, where here flipping a cluster means that for all sites in the cluster
the spins are exchanged. For example, the cluster is flipped with the Metropolis-type probability
\begin{equation}
 p_{exc,C} = \mbox{min}  [1, \exp(-2 \sum_{x \in C} h_{x,embed}) ] \;,
\end{equation}
where the sum runs over all sites $x$ that belong to the given cluster $C$.

Here we study two films of the thicknesses $L_{0,1} = L_0+d/2$ and $L_{0,2} = L_0-d/2$, 
where $d=1$, $2$, $... \;\;$. In the case of system 1, the spins at $x_0=0$ and $L_{0,1}+1$ 
are fixed in order to implement the boundary conditions, while for system 2, the 
spins at $x_0=0$ and $L_{0,2}+1$ are fixed. In order to have the same number of sites
for both systems 1 and 2, we add in the case of system 2 auxiliary spins at 
$x_0 = L_{0,2} + 2, ..., L_{0,1}+1$, which assume the same value as those at $x_0=L_{0,2} +1$. 

Clusters are constructed according to the delete probability given by eq.~(\ref{pdsimple}). 
This means that a link between a pair of neighbor sites is frozen with the probability 
$p_f = 1 -p_d$. Two sites belong to the same cluster, if there exists a chain of frozen links
that connects the two sites.  In order to keep the boundary conditions in place, only clusters 
are flipped that do not contain sites with fixed spins.

The purpose of the exchange cluster algorithm is to obtain a variance reduced estimator of 
$\langle \Delta E \rangle$. To this end, it is optimal to exchange as many spins as possible. Hence only those
spins are not exchanged that belong to clusters that contain fixed spins.  To this end we 
have to construct only those clusters that contain fixed spins. Starting the cluster at 
$x_0=0$,  the cluster can not grow to $x_0=1$, since $s_{(0,x_1,x_2),1} = s_{(0,x_1,x_2),2}$
and hence $\beta_{embed} =0$, which implies that $p_d = 1$.  Only starting from $x_0=L_{0,2}+1$,
a cluster containing fixed spins of system 2 only, might grow to $x_0=L_{0,2}$. 
Hence we start the construction of the frozen clusters by running through all sites 
$x=(L_{0,2}+1,x_1,x_2)$ and add the site $y=(L_{0,2},x_1,x_2)$ to the frozen clusters with the 
probability $p_f=1-p_d$, eq.~(\ref{pdsimple}). Note that in this initial step we have to check only 
this single neighbor, since the other ones are frozen anyway. Then the construction of the 
frozen clusters is completed using a standard algorithm for the cluster search. 

In our C-program the spins are stored in an array
\verb|char spins[I_D][L_Z][L][L];|
where \verb|I_D| equals two and \verb|L_Z| equals $L_0+d/2+2$. 
Similar to the case of the plate-sphere geometry, it turns out that the frozen clusters
usually take only a small fraction of the lattice.
Therefore, in order to save CPU time we do not copy all spins
outside the frozen clusters from \verb|spins[0][][][]|  to \verb|spins[1][][][]| and
vice versa. Instead, we do that for the spins that belong to frozen clusters.
This way, the systems $1$ and $2$ 
interchange their position in the array \verb|spins|. In order to keep track of where
the systems are stored in the array \verb|spins|, we introduce the array
\verb|int posi[I_D];| where the index \verb|i_d| equals $1$ or $2$ and
\verb|posi[i_d]| indicates whether system $1$ is stored in \verb|spins[0][][][]| 
or \verb|spins[1][][][]| and system $2$ correspondingly. 
Implemented this way, the CPU-time required by the cluster exchange update is essentially 
proportional to the size of the frozen clusters.

\subsection{Construction of improved differences}
The main purpose of the exchange cluster is to allow us to define improved estimators
for the difference of observables defined in systems 1 and 2. Here this is mainly $\Delta E$, 
however also other quantities can be computed efficiently as we shall see below. 
The basic idea behind these improved differences is that large parts of the configurations
are swapped between the two systems. This way we get exact cancellations
for most of the lattice volume. Let us
consider an observable $A$ that is defined for both systems $1$ and $2$. We are aiming at 
a variance reduced estimator for the difference 
\begin{equation}
\Delta A = A_1-A_2 \; .
\end{equation}
To this end we make use of the correlation of the configuration of system 1 at Markov-time $t+1$
with that of system 2 at Markov-time $t$, and vice versa:
\begin{equation}
\label{improved_general}
 \Delta A_{imp}  =  \frac{1}{2} \left([A_{1,t} - A_{2,t+1}] + [A_{1,t+1} - A_{2,t}]\right) \;,
\end{equation}
where the second index of $A$ now gives the position of the configuration in the Markov chain
and $t$ and $t+1$ are separated by a single exchange cluster update.

Let us work out eq.~(\ref{improved_general}) explicitly for $\Delta E$:    
\begin{eqnarray}
\label{improvedE}
\Delta E_{imp} &=& \frac{1}{2}
 \sum_{<xy>}\left( [s_{x,2}^{(t)} s_{y,2}^{(t)} - s_{x,1}^{(t)}  s_{y,1}^{(t)}] +
                    [s_{x,2}^{(t+1)} s_{y,2}^{(t+1)} - s_{x,1}^{(t+1)} s_{y,1}^{(t+1)}] \right)
 \nonumber \\
 &=& \frac{1}{2} \sum_{<xy>} \left( [s_{x,2}^{(t)} s_{y,2}^{(t)} - s_{x,1}^{(t+1)} s_{y,1}^{(t+1)}] +
                                 [s_{x,2}^{(t+1)} s_{y,2}^{(t+1)} - s_{x,1}^{(t)} s_{y,1}^{(t)}] \right)
 \nonumber \\
&=& \frac{1}{2} \sum_{<xy> \in C_f}
    \left( [s_{x,2}^{(t)} s_{y,2}^{(t)} - s_{x,1}^{(t+1)}  s_{y,1}^{(t+1)} ] +
              [s_{x,2}^{(t+1)} s_{y,2}^{(t+1)} - s_{x,1}^{(t)} s_{y,1}^{(t)}] \right)
\end{eqnarray}
where $<xy> \in C_f$ means that at least one of the sites $x$ or $y$ belongs to a frozen
cluster. Hence also the numerical effort to compute $\Delta E_{imp}$ is approximately proportional 
to the size of the frozen clusters.
Note that for our choice of the update
\begin{equation}
 s_{x,1}^{(t+1)} s_{y,1}^{(t+1)} = s_{x,2}^{(t)} s_{y,2}^{(t)} \;\;\; \mbox{and} \;\;\;
 s_{x,2}^{(t+1)} s_{y,2}^{(t+1)} = s_{x,1}^{(t)} s_{y,1}^{(t)}
\end{equation}
for all nearest neighbor pairs $<x,y>$ where neither $x$ nor $y$ belongs to a frozen
cluster.

\subsection{The simulation algorithm, benchmarks and tuning of parameters}
The exchange cluster algorithm on its own is not ergodic, since it keeps
the total number of spins of a given value fixed. Therefore  we performed
in addition updates of the individual systems, using standard cluster 
and local updates \cite{BrTa89}. 
In all our simulations we used the
Mersenne twister algorithm \cite{twister} as pseudo-random number generator.

\subsubsection{cluster algorithm for the individual system}
We used the standard delete probability $p_d = \mbox{min}[1,\exp(-2 \beta s_x s_y)]$ 
in the construction of the clusters.  One has to take into account that
clusters that contain sites with fixed spins can not be flipped. Flipped means that
all spins that belong to the cluster are multiplied by $-1$. 
We have used two types of 
cluster-updates. In the first one, denoted by SW-cluster algorithm in the following,
we flip the clusters, that do not contain fixed sites, following  ref. \cite{SwWa87}, 
with the probability $1/2$. In the second one, denoted by B-cluster algorithm in the 
following, clusters that do not contain fixed sites, are always flipped. This has the 
technical advantage, that actually only clusters that contain sites with fixed
spins have to be constructed. All other spins are flipped. For $(O,O)$ 
boundary conditions, only the  SW-cluster algorithm is used, since for $s_x=0$ or
$s_y=0$ we get $p_d=1$ and hence there are no clusters that contain both sites of the
interior and the boundary.

\subsubsection{Todo-Suwa algorithm}
The authors of \cite{ToSu13} have pointed out that auto-correlation times of 
local updating algorithms can be reduced by a significant factor, when one 
abstains from detailed balance and only demands the sufficient condition of
balance. This idea still leaves considerable freedom for the design of the 
algorithm. Todo and Suwa suggest to order the possible values of the local
spin on a cycle. Then one preferentially updates  in one of the two directions
on the cycle. For the precise description see ref. \cite{ToSu13}. 
Todo and Suwa have tested their algorithm for example
at the 4- and 8-state Potts model in two dimensions in the neighborhood of
the critical point. They find a reduction of the auto-correlation time 
compared with the heat-bath algorithm by a factor of 2.7 and 2.6 for the 
4- and 8-state Potts model, respectively.  In the case of the improved Blume-Capel
model on the simple cubic lattice at the critical point one finds a reduction 
by a factor of about 1.7 compared with the heat-bath algorithm \cite{Gutsch}.
Since we failed to find a prove of ergodicity for the Todo-Suwa local update,
sweeping through the lattice in type-writer fashion, we performed heat-bath
sweeps in addition. Note that for the heat-bath the prove of ergodicity is 
trivial.

\subsubsection{The update cycle}
\label{updatecycle}
We initialized the spins that are not fixed by choosing one of the three
possible values with equal probability. Then we equilibrated the systems
by performing 1000 update cycles consisting of one heat-bath sweep, one 
SW-cluster update, one Todo-Suwa sweep and one B-cluster update. In the
case of $(O,O)$ boundary conditions, the B-cluster update is omitted.

After this initial phase of the simulation we added $n_{exc}$ exchange cluster updates
to each update cycle. Furthermore, since the frozen exchange clusters are very much localized 
at the boundary, we performed for each exchange cluster update a local update with the 
Todo-Suwa algorithm of the $i_r$ layers of the lattices that are closest to the
upper boundary. Only in a few preliminary tests we shall use a different sequence of 
updates, which will be stated below.

\subsubsection{Tuning the parameters of the update cycle and benchmarking the algorithm}
\label{tuning}
First we tested the performance of the exchange cluster algorithm 
for $(+,-)$ boundary conditions at the critical point $\beta_c = 0.387721735$.
To keep things simple, we first used the following update sequence:
A global sweep with the heat-bath algorithm over both systems followed by 
one exchange cluster update, combined with a random translation of one system 
with respect to the other in the transversal directions.  Our results
are summarized in table \ref{performance}. In all cases $10^5$ update cycles and 
measurements were performed. In the third column we give the 
size of the frozen exchange clusters per area $S_c$. The 
$d=L_{0,1}-L_{0,2}$ layers, where the spins of system $2$ are fixed and those
of system $1$ are not, are taken into account in $S_c$. This means that $S_c$
at least equals to $d$. We find that $S_c$ is small compared with the thickness
of the films in all cases. For given $d$ it depends very little on the thickness $L_0$.
As one might expect, it increases with increasing $d$.  We give the variance 
of $\Delta E_{imp}$ and of the energies $E_1$ and $E_2$ normalized by the area $L^2$, 
since this normalized number should  have a finite $L \rightarrow \infty$ limit.
We find that the variance of
$\Delta E_{imp}$ is reduced compared with the sum of the variances 
of the energies $E_1$ and $E_2$ of the individual systems. For fixed $d$, the ratio 
of the two variances increases with increasing lattice size.  On the other hand, 
the advantage of the improved estimator becomes smaller with increasing $d$.
Often variance reduced estimators have a larger integrated auto-correlation time 
than the basic quantity. 
Here, in  contrast we observe that the integrated auto-correlation time of
$\Delta E_{imp}$ is considerably smaller than those of the energies $E_1$ and $E_2$ of
the individual systems.

\begin{table}
\caption{\sl \label{performance}
We study the properties of the exchange cluster algorithm for $(+,-)$ boundary 
conditions at $\beta_c$. The transversal extension of the lattices is $L=32$, 
$64$, and $128$ for $L_{0,2}=8$, $16$, and $32$, respectively. For the 
definition of the quantities and a discussion see the text.
}
\begin{center}
\begin{tabular}{rrcclcll}
\hline
 $L_{0,1}$ & $L_{0,2}$ & $S_c$ & var $(\Delta E_{imp})/L^2$ & $\tau_{int,imp}$ 
    &  $[$var$(E_1)+$var$(E_2)]/L^2$ & $\tau_{int,E_1}$ & $\tau_{int,E_2}$ \\
\hline
  9  &  8 &1.4462(11) & 35.5(2)&1.21(2) & \phantom{0}144.2(6) & 1.67(3)  
&   1.37(3) \\
 17  &  16 &1.4538(12) & 58.4(3)&1.40(4) & \phantom{0}378.(2.) & 2.97(10) &
  2.62(9) \\
 18  &  16 &2.7966(21) & 96.3(6)&2.28(7) & \phantom{0}399.(2.) & 3.61(13) & 
 2.85(10) \\
 33  &  32 &1.4547(11) & 89.8(5)& 1.52(6)& \phantom{0}995.(6.) & 9.6(7)  &  
9.2(6)  \\
 34  &  32 &2.8001(22) &152.4(10)&2.67(11)&\phantom{0}982.(6.) & 9.6(7) &  
7.5(5) \\
 36  &  32 &5.3388(41) &252.(2.)\phantom{0}&5.70(33) &1046.(7.)& 9.6(7) & 8.3(6) \\
\hline
\end{tabular}
\end{center}
\end{table}

Next we studied an update cycle that includes cluster updates of the individual
films. In particular we used the update cycle stated in section \ref{updatecycle}
above: one sweep with the heat-bath algorithm, 
a SW-cluster update, one sweep with the Todo-Suwa algorithm and a B-cluster update.

Motivated by the fact that $S_c$ is small and hence the CPU-time required by the exchange cluster
update is little and that the integrated autocorrelation time $\tau_{int,imp}$ is relatively
small, we performed $n_{exc}$ exchange cluster updates for each update cycle. Furthermore, 
since the frozen exchange clusters are very much localized at the upper boundary, 
a sweep with the local Todo-Suwa algorithm of the $i_r$ layers that are closest to 
the upper boundary is performed. 
In the following we try to find the optimal choice for the parameters $n_{exc}$ and $i_r$. 
Again we perform this study at the critical point for $(+,-)$ boundary conditions.
 
As example let us consider the pair of lattices characterized by $d=1$, $L_0=32.5$ and $L=128$.  
On our CPU, the time required by a single exchange cluster update is about $0.014$ times 
the one needed for the total of the SW-cluster, B-cluster updates and the heat-bath and Todo-Suwa
sweeps. Updating one layer in both lattices using the Todo-Suwa algorithm takes about $0.0049$ times
the CPU-time of these updates. Hence the CPU-time required by the complete cycle is proportional to
\begin{equation}
 t_{mix} = 1 + n_{exc} (0.014 + 0.0049 i_r) \;\;.
\end{equation}

\begin{figure}
\begin{center}
\includegraphics[width=14.5cm]{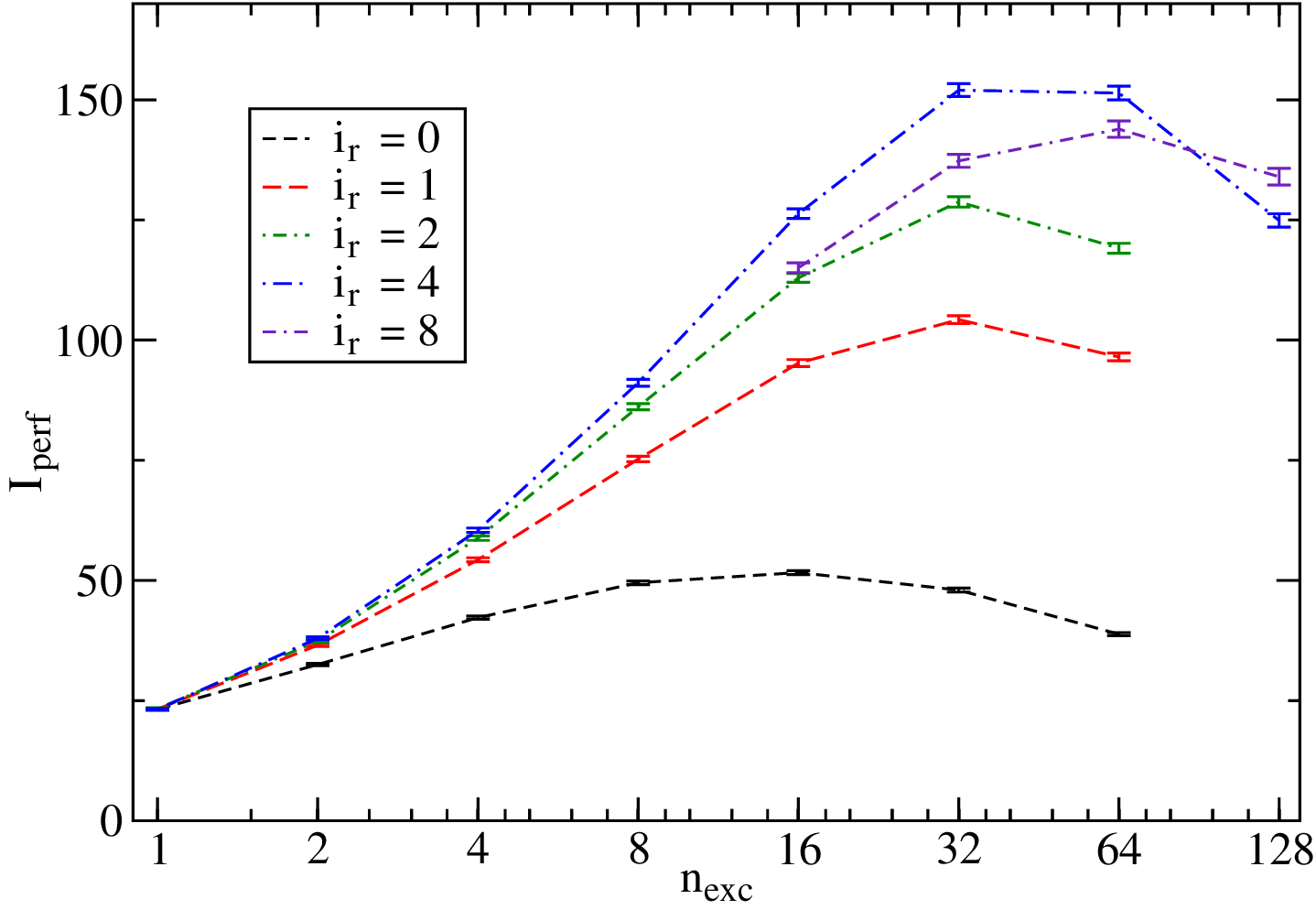}
\caption{\label{perfplot}
We study $(+,-)$ boundary conditions at the critical point. We simulated a pair of lattices
characterized by $L_0=32.5$, $d=1$ and $L=128$.
We plot the performance index $I_{perf}$ defined in eq.~(\ref{iperf}) as a function 
of the number $n_{exc}$ of exchange cluster updates per update cycle. Results
are given for $i_r=0$, $1$, $2$, $4$, and $8$. For a discussion see the text.
}
\end{center}
\end{figure}

We define a performance index  as 
\begin{equation}
\label{iperf}
I_{perf} = \frac{\mbox{var}[E_1-E_2] \; \tau_{int,E_1-E_2}} 
                {t_{mix} \; \mbox{var}[\Delta E_{imp}] \; \tau_{int,imp} } \;\;,
\end{equation}
where $\mbox{var}[E_1-E_2]$ and $\tau_{int,E_1-E_2}$ are taken from a simulation with $n_{exc}=0$, 
i.e. without any exchange cluster update. We simulated for a large number of values of $n_{exc}$ and 
$i_r$. The number of update cycles ranges from   $2 \times 10^5$ to $10^6$. Our results are plotted
in Fig. \ref{perfplot}. Among our choices, the optimal performance is reached for $i_r=4$ and 
$n_{exc} = 32$. For these parameters the improvement is $I_{perf}=152.0(1.3)$, which means that 
the improved cluster exchange estimator allows to reduce the statistical error
by more than a factor of 12 at a given CPU-time. 
 We also see that this maximum is rather shallow, which means that no accurate 
fine-tuning of the algorithm is needed to reach a fair fraction of the optimum. 

We performed an analogous study for $L_0=16.5$ and $64.5$, simulating a smaller
number of values
of $i_r$ and $n_{exc}$, focussing on finding the optimal values.  For $L_0=16.5$
the maximum is also reached for  $i_r=4$ and $n_{exc} = 32$ with $I_{perf}=45.5(3)$.
Also here the maximum of $I_{perf}$ is very shallow. For example for 
$i_r=4$ and $n_{exc} = 16$ we get $I_{perf}=40.3(3)$ or for $i_r=2$ and 
$n_{exc} = 32$ we get  $I_{perf}=43.0(3)$. For $L_0=64.5$ the optimum is located at
$i_r=8$ and $n_{exc} = 64$ with $I_{perf}=553.(13.)$. For $i_r=4$ and $n_{exc}=32$
we get $I_{perf}=505.(10.)$.  For $d=1$ fixed, $I_{perf}$ increases almost
like $L_0^2$ with increasing thickness.  This means that the problem of the increasing 
variance, eq.~(\ref{problemvar}), of the standard estimator is cured by the improved
estimator.

Here we performed a random translation of the systems with respect to each other in the 
lateral directions
performing the cluster exchange update. Studying for example random disorder at
the boundary, this symmetry is not available. Therefore we checked how much the 
performance gain $I_{perf}$ depends on these translations. To this end we repeated
the simulations for $L_0=32.5$, $i_r=4$ and $n_{exc} = 32$ without these 
translations. It turns out that $I_{perf}$ is smaller by a factor of about $1.6$.
This means that one certainly should use the translation when the symmetry is 
present. However the effectiveness of the cluster exchange update does not crucially
depend on it. 

Likely further improvements can be achieved by exploiting for example reflection
symmetries. Also a more elaborate update cycle might improve the performance. We did not 
further explore these ideas. 
Actually we did not systematically tune the parameters $i_r$ and $n_{exc}$ for the whole
range of temperatures and different boundary conditions  discussed below.
Throughout we used $n_{exc}=20$. In fact, we had started our simulations before performing
the systematic tuning discussed above. 
\section{Thermodynamic Casimir force for strongly symmetry breaking boundary conditions}
\label{strongsection}

These boundary conditions have been studied by using 
Monte Carlo simulations of the Ising model \cite{VaGaMaDi07,VaGaMaDi08,VaMaDi11,MHcorrections}
and the improved Blume-Capel model \cite{MHstrong,MHcorrections} before.

Here we simulated films of the thicknesses $L_0=16.5$, $32.5$, and $64.5$.  Throughout we use $d=1$. 
In the case of $(+,+)$ boundary conditions the correlation length of the film stays small, it reaches
a maximum at $x = t [L_{0,eff}/\xi_0]^{1/\nu}  \approx 7$, 
where $\xi_{2nd,film} \approx 0.145 L_{0,eff}$, see section VII B of 
ref. \cite{MHstrong}.  We simulated lattices of the transversal linear size $L=64$ and $128$ for 
$L_0=16.5$, $L=128$ and $256$ for $L_0=32.5$ and $L=256$  for $L_0=64.5$. Given the relatively small
correlation length of the film, these transversal extensions should clearly be sufficient to keep
finite $L$ effects at a negligible level. This is explicitly verified by the comparison of results
obtained for the two different values of $L$ simulated for $L_0=16.5$ and $32.5$.
In the case of $(+,-)$ boundary conditions the correlation length of the film is monotonically 
increasing with increasing inverse temperature $\beta$. The physical origin of this behavior are
fluctuations of the interface between the two phases that arises in the low temperature phase.
At the critical point $\xi_{2nd,film} \approx 0.212 L_{0,eff}$ \cite{MHstrong}. Results for the
full range of $x$ that we have studied are given in Fig. 7 of \cite{MHstrong}.  Here, in order
to keep finite $L$ effects negligible, we have chosen $L \gtrapprox 10 \xi_{2nd,film}$.  The 
largest values of $L$ that we simulated are $L=512$, $1024$ and $1024$ for $L_0=16.5$, $32.5$, 
and $64.5$, respectively.

For both $(+,+)$ and $(+,-)$ boundary conditions, we took $i_r=2$, $4$ and $8$ for $L_0=16.5$, $32.5$, and $64.5$, 
respectively. As already mentioned above, we have chosen $n_{exc}=20$ for all our simulations.
As discussed above in section \ref{tuning}, in particular for $L_0=64.5$ a larger value of 
$n_{exc}$ would have been a better choice.

In most of the simulations we performed $10^5$ update cycles. Only for $(+,-)$ for 
$(L_0,L) =(32.5,1024)$, $(64.5,512)$, and $(64.5,1024)$ we performed less update cycles, where
the minimal number was $29300$. In total we used about $1.5$ and $3.5$ years 
of CPU time on a single core of an AMD Opteron 2378 for $(+,+)$ and $(+,-)$
boundary conditions, respectively.

\begin{figure}
\begin{center}
\includegraphics[width=14.5cm]{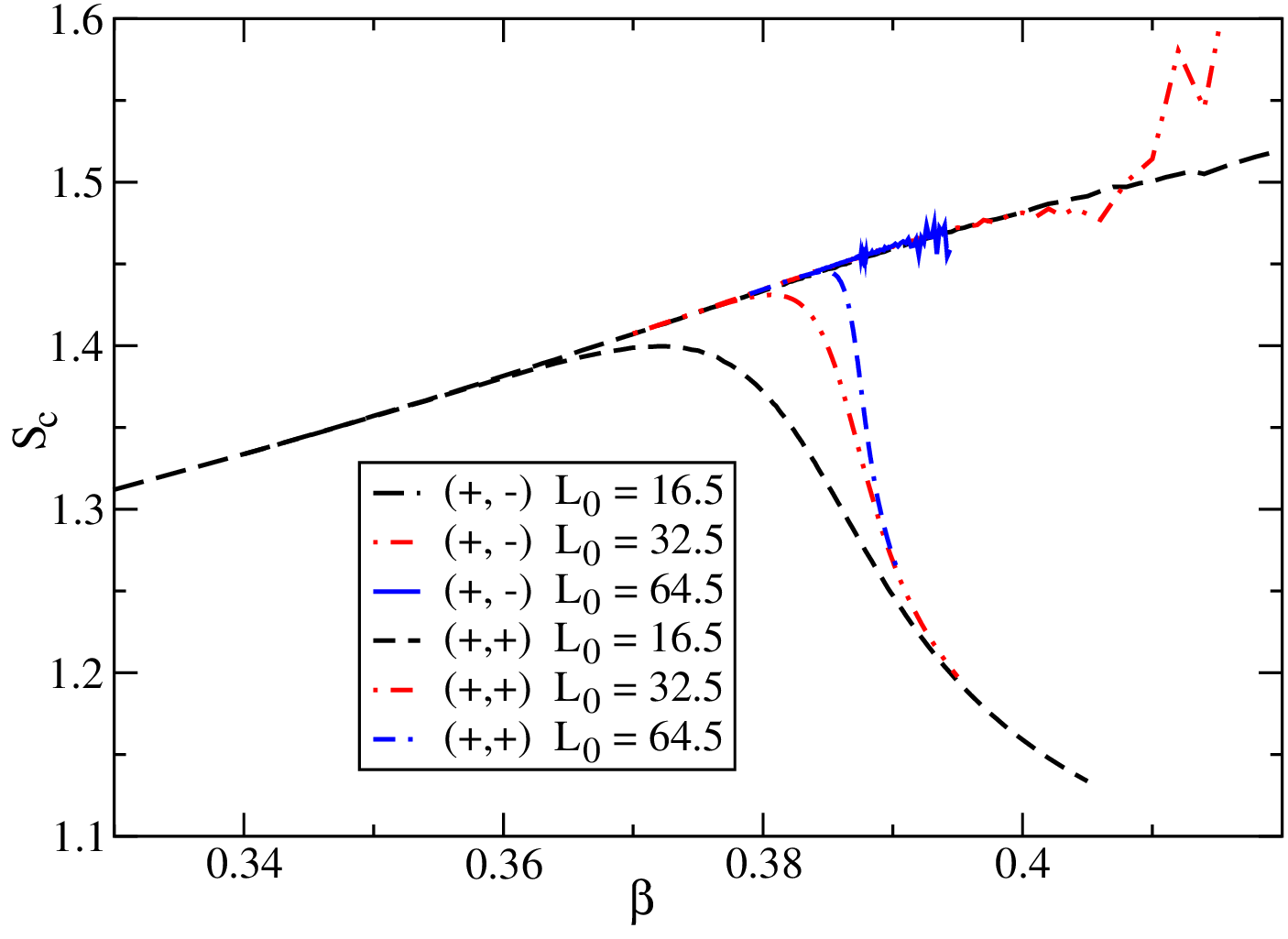}
\caption{\label{ballall}
The average size $S_c$ of the frozen exchange clusters per area  is plotted as a function of $\beta$. We give
results for the thicknesses $L_0=16.5$, $32.5$, and $64.5$ for $(+,+)$ and $(+,-)$ boundary 
conditions.
}
\end{center}
\end{figure}
Before going to the physics results, let us discuss the properties of the exchange cluster
algorithm.
In Fig. \ref{ballall} we plot the average size $S_c$ per area of the frozen exchange clusters as
a function of $\beta$. For small values of $\beta$, the curves for both types of 
boundary conditions as well as all three thicknesses of the film fall on top of each other.
For small $\beta$, $S_c$ slowly increases with increasing $\beta$. In the case of 
$(+,-)$ boundary conditions $S_c$ increases, up to statistical fluctuations, in the whole
range of $\beta$ that we have studied. In the neighborhood of $\beta_c$ 
no particular change of the behavior can be observed. In Fig. \ref{ballall} we give no error bars, 
in order to keep the figure readable. We have convinced ourself that the fluctuations that 
can be seen for $(+,-)$ boundary conditions for $L_0=32.5$ and $64.5$ at large values of $\beta$
can be explained 
by large statistical errors due to large auto-correlation times. These are likely caused by slow
fluctuations of the interface between the phases of opposite magnetization. The analogue problem 
for anti-periodic boundary conditions is discussed in ref. \cite{Meyer}. Here we made no attempt 
to adapt the special cluster algorithm of ref. \cite{Meyer} to $(+,-)$ boundary conditions.

In the case of $(+,+)$ boundary conditions, starting from a certain value of $\beta$ that 
depends on the thickness $L_0$, $S_c$ departs from the curve for $(+,-)$ boundary conditions.
At the resolution of our plot, this happens when the bulk correlation length becomes $\xi \approx L_0/7$. 
At some $\beta(L_0) < \beta_c$, $S_c$ reaches a maximum. In the low temperature phase, as 
$\beta$ increases, again the curves for different $L_0$ fall on top of each other. 

With respect to the performance of the exchange cluster algorithm it is important to 
note that in all cases $S_c$ remains small compared with the thickness $L_0$ in the whole range of 
$\beta$ that we have studied.

Next we discuss how much the statistical error is reduced by employing the improved
estimator of the energy difference.  Here we can not use $I_{perf}$ defined in eq.~(\ref{iperf}), 
since we did not perform simulations with $n_{exc}=0$ for the whole range of $\beta$.
Hence we study the ratio 
\begin{equation}
\label{gaindef}
 \mbox{gain} = \frac{\epsilon(\Delta E)}{\epsilon(\Delta E_{imp})}
\end{equation}
where $\epsilon(\Delta E)$ and $\epsilon(\Delta E_{imp})$ are the statistical errors of the 
energy difference computed in the standard and the improved way, respectively. In the case
of the standard estimator
we have computed $\epsilon^2(\Delta E(L_0)) = \epsilon^2(E(L_0+1/2)) + \epsilon^2(E(L_0-1/2))$
naively, not taking into account the statistical correlation of the two quantities due to the 
exchange cluster updates. Note that eq.~(\ref{gaindef}) gives a ratio of statistical errors.
Hence this gain has to be squared to be compared with $I_{perf}$ defined in eq.~(\ref{iperf}). 

\begin{figure}
\begin{center}
\includegraphics[width=14.5cm]{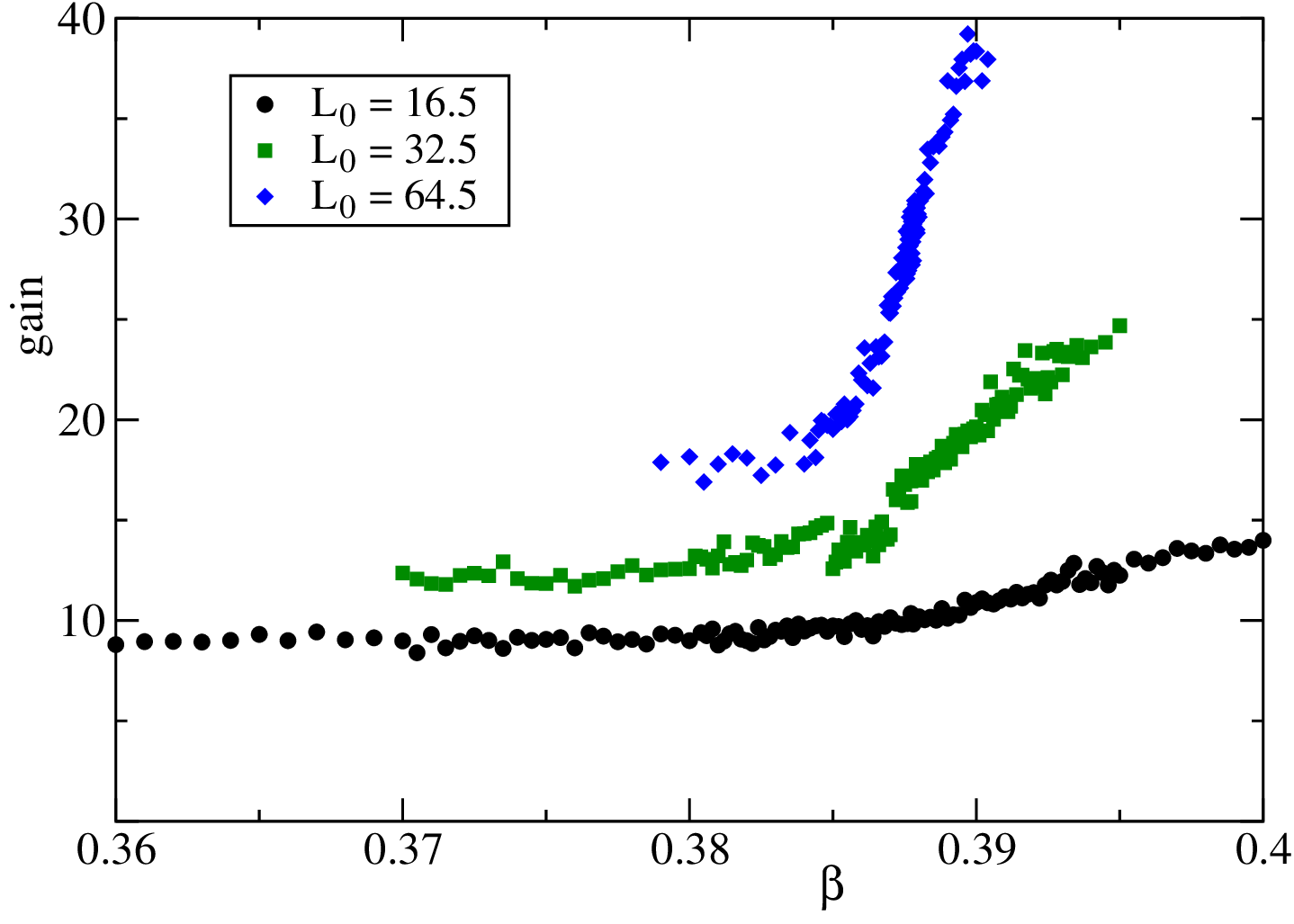}
\caption{\label{plotgainpara}
We plot the quantity ``gain'' defined in eq.~(\ref{gaindef}) as a function of the inverse temperature $\beta$
for $(+,+)$ boundary conditions and the thicknesses $L_0=16.5$, $32.5$ and $64.5$.
}
\end{center}
\end{figure}

In figure \ref{plotgainpara} this gain is plotted for $(+,+)$ boundary conditions. 
At small values of $\beta$, the gain 
depends very little on $\beta$.  At $\beta$ slightly smaller than $\beta_c$ the gain starts to 
increase with $\beta$. At larger values of $\beta$ the gain increases approximately linearly
with $\beta$. It is interesting to note that the gain increases with increasing thickness of 
the lattice size. At $\beta_c$ we get gain $\approx 10.2$, $17.3$, and $28.5$ 
for $L_0=16.5$, $32.5$ and $64.5$, respectively.

For $(+,-)$ boundary conditions we find that the gain depends only weakly 
on the inverse temperature $\beta$. 
At $\beta_c$ we get  gain $\approx 8.2$, $12.5$, and $15.2$
for $L_0=16.5$, $32.5$ and $64.5$, respectively.   This means that we 
profit less from the cluster exchange estimator than in the case of $(+,+)$ 
boundary conditions. The square of gain is quite roughly equal to 
$I_{perf}$ determined in the section above. 

Now let us turn to the analysis of our numerical results for the 
thermodynamic Casimir force.
Following refs. \cite{MHstrong,MHcorrections} we chose the starting point 
$\beta_0$ of the integration~(\ref{integration}) such that the 
approximation discussed in sec. IV A of ref. \cite{MHstrong} is still valid. 
We get
\begin{equation}
\label{startingpoint}
\Delta f_{ex}(\beta_0) = \pm \frac{C^2(\beta_0)}{\xi^2(\beta_0)}
\frac{\exp[-(L_0 + 1+d/2)/\xi(\beta_0)] -\exp[-(L_0 + 1-d/2)/\xi(\beta_0)]}{d}
\;,
\end{equation}
where we have  $+$ for $(+,+)$ boundary conditions and $-$
for $(+,-)$ boundary conditions. The numerical values of $C^2(\beta_0)$ and
$\xi(\beta_0)$ are taken from ref. \cite{MHcorrections}. By comparing
results obtained with different choices of $\beta_0$ we found that the
approximation~(\ref{startingpoint}) is accurate at the level of our statistical
error for $L_0/\xi(\beta_0) \gtrapprox 8$. To be on the safe side,
we used $L_0/\xi(\beta_0) > 10$ in the following.

\begin{figure}
\begin{center}
\includegraphics[width=14.5cm]{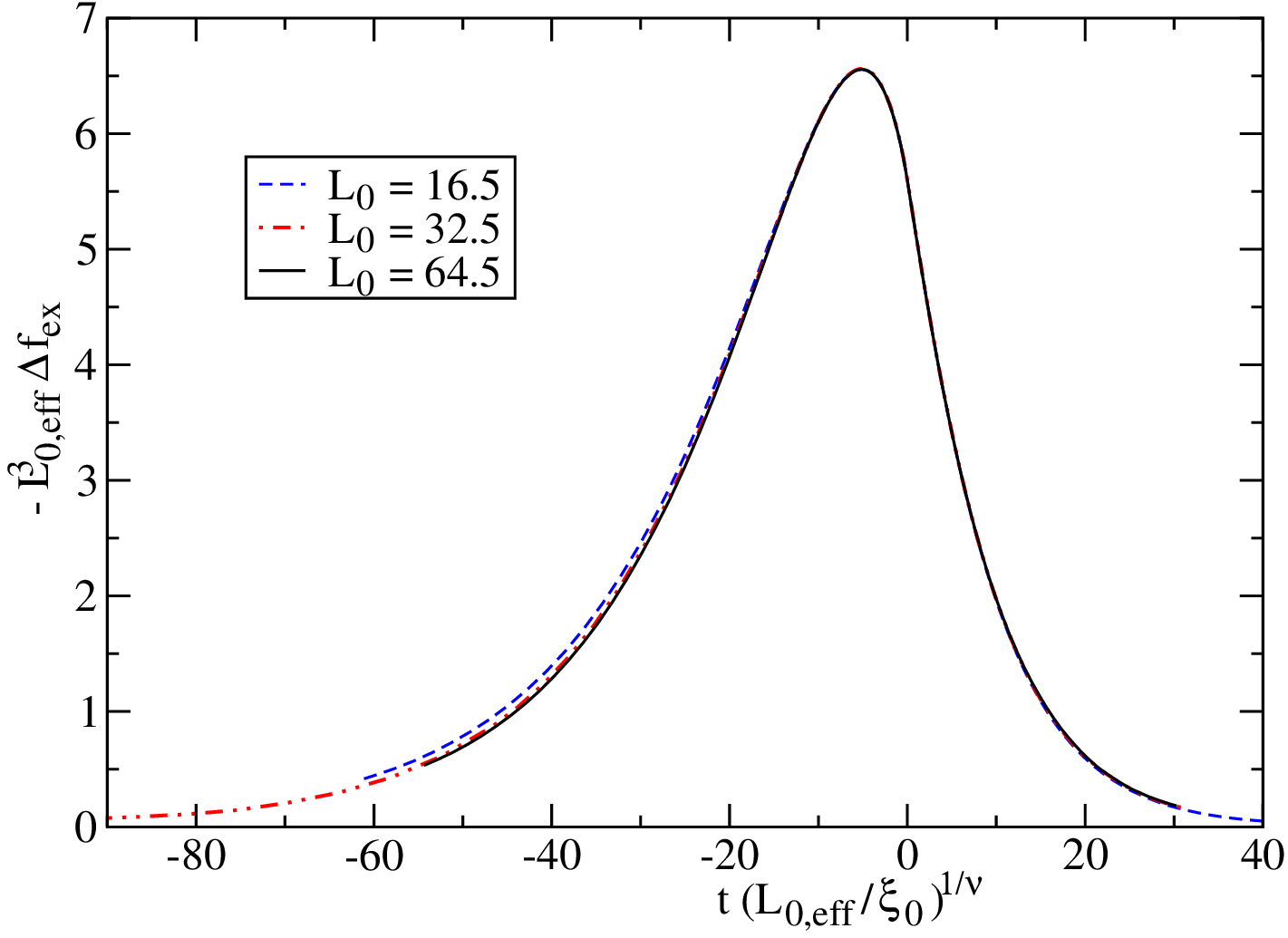}
\caption{\label{antiplot}
Numerical results for the scaling function $\theta(x)$ for $(+,-)$ boundary conditions.
We plot $-L_{0,eff}^3 \Delta f_{ex}$ as a function of $t (L_{0,eff}/\xi_0)^{1/\nu}$, where
$L_{0,eff} = L_0 +L_s$ with $L_s=1.91$, $\xi_0=0.2283$, and $\nu=0.63002$. The thicknesses
of the film are $L_0 = 16.5$, $32.5$, and $64.5$. The error bars are typically smaller than 
the thickness of the lines. 
}
\end{center}
\end{figure}

Let us discuss the results obtained for the scaling function 
$\theta(x) \simeq -L_{0,eff}^3 \Delta f_{ex}$, where $x=t [L_{0,eff}/\xi_0]^{1/\nu}$. 
In Fig. \ref{antiplot} we give our results for $(+,-)$ boundary conditions.
For $x \gtrapprox -15$ the curves for the 
three different thicknesses fall nicely on top of each other. For $x \lessapprox -15$ 
we see a small deviation of the result for $L_0=16.5$ from the other two thicknesses.
The difference between $L_0=32.5$ and $64.5$ can hardly be resolved.
Hence we are confident that corrections to scaling are well under control and the 
numerically important contributions are well described by the effective thickness 
$L_{0,eff} = L_0 + L_s$ with $L_s=1.91(5)$. Finally let us discuss the maximum 
of $\theta_{(+,-)}$. Via the zero of $\Delta E_{ex}$ we find $\beta_{max} = 0.392560(10)$, 
$0.389512(5)$, and $0.388355(3)$ for $L_0= 16.5$, $32.5$, and $64.5$, respectively.
This corresponds to $x_{max}=t_{max} [(L_0 +L_s)/\xi_0]^{1/\nu} =-5.139(11)[22]$, $-5.131(14)[12]$,
and $-5.154(24)[6]$, where the number in $[]$ gives the error due to the 
uncertainty of $L_s$. Note that the dependence on $\nu$ essentially cancels when taking into
account the dependence of the estimate of $\xi_0$ on $\nu$, eq.~(\ref{xi0}). 
The maximal value of $-L_{0,eff}^3 \Delta f_{ex}$ is $6.558(3)[54]$, $6.561(3)[29]$ and 
$6.556(7)[15]$, where again the number in $[]$ gives the error due to the uncertainty of $L_s$.
The results obtained for the different thicknesses nicely agree. We conclude
\begin{equation}
 x_{max} =-5.14(4) \;\;,\;\;\;  \theta_{(+,-)}(x_{max}) = 6.56(3) \;.
\end{equation}
These estimates are fully consistent with those of our previous work \cite{MHcorrections}. 
Note that the error bars of the final estimates are not reduced compared with \cite{MHcorrections}.
This is mainly due to the fact that the same estimate of $L_s$ is used and that the 
uncertainty of $L_s$ is a major source of the error.  

For a comparison of the result for $\theta_{(+,-)}(x)$ given in \cite{MHcorrections}, which 
is fully consistent with the present result, with the results of Monte Carlo simulations 
of the Ising model \cite{VaGaMaDi08},
experiments on a binary liquid mixture \cite{FuYaPe05} and the extended de Gennes-Fisher 
local-functional method see Fig. 1  of ref. \cite{UpBo13}.

\begin{figure}
\begin{center}
\includegraphics[width=14.5cm]{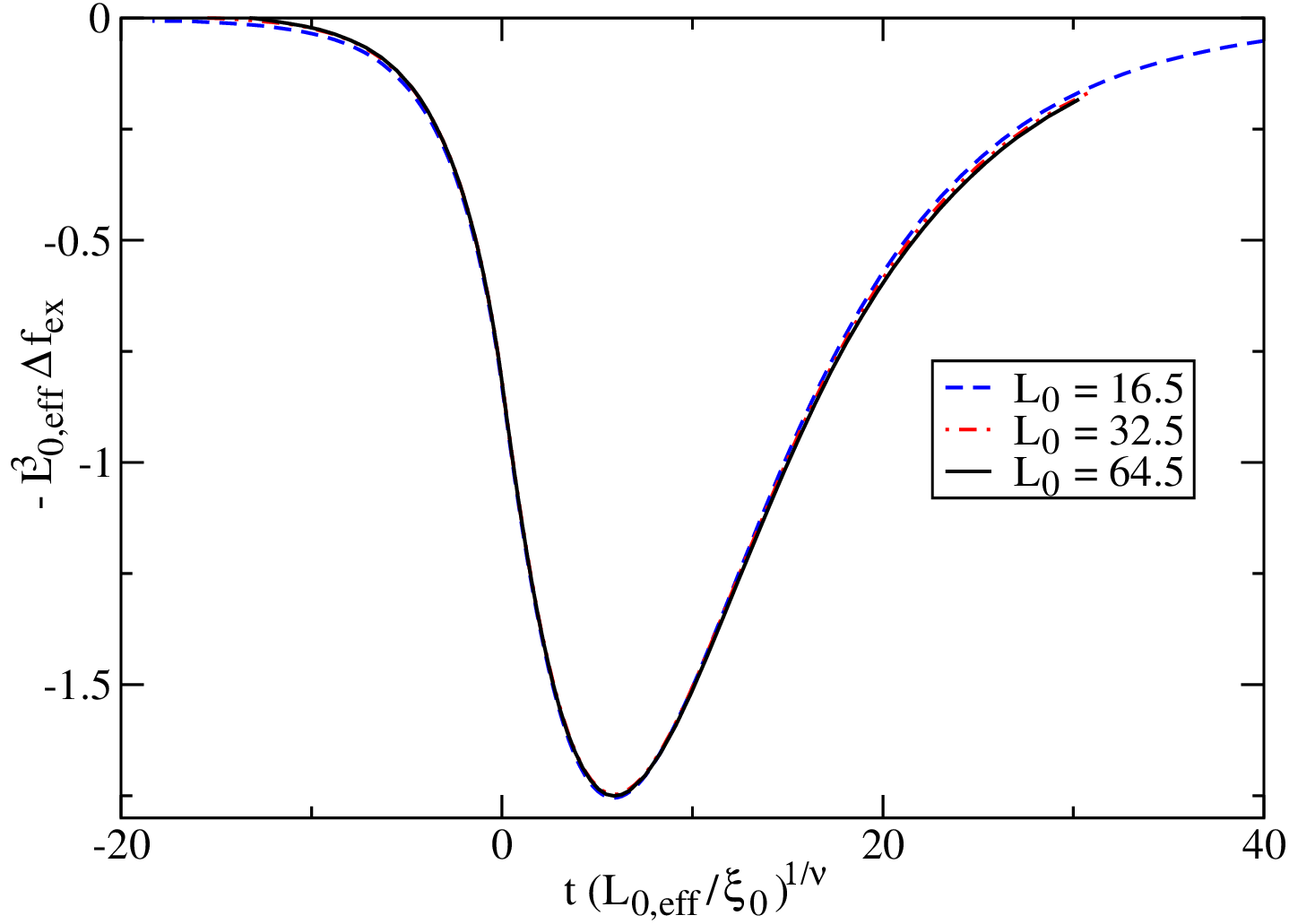}
\caption{\label{paraplot}
Same as previous figure, but for $(+,+)$ instead of $(+,-)$ boundary conditions.
}
\end{center}
\end{figure}

In Fig. \ref{paraplot} we give our numerical results for $\theta_{(+,+)}(x)$.  In the neighborhood
of the minimum of $-L_{0,eff}^3 \Delta f_{ex}$ the curves for the three different thicknesses
fall nicely on top of each other. But also for small and large values of the scaling variable $x$
the differences remain small. In particular the curves for $L_0=32.5$ and $64.5$ can hardly be
discriminated. We conclude that similar to the case of $(+,-)$ boundary conditions, corrections 
to scaling are well under control. Let us look at the minimum of $\theta_{(+,+)}$ in more detail.
We find $\beta_{min} = 0.382213(22)$, $0.385670(10)$ and $0.387001(7)$ for $L_0= 16.5$,
for $L_0= 16.5$, $32.5$, and $64.5$, respectively. This corresponds to $x_{min}=
t_{min} [(L_0 +L_s)/\xi_0]^{1/\nu} = 5.851(23)[25]$, $5.881(29)[14]$, and $5.866(57)[7]$,
where the number in $[]$ gives the error due to the uncertainty of $L_s$.
The minimal value of $-L_{0,eff}^3 \Delta f_{ex}$ is $-1.755(3)[14]$, $-1.747(4)[8]$, and  
$-1.750(7)[4]$, where again the number in $[]$ gives the error due to the uncertainty of $L_s$.
We conclude
\begin{equation}
 x_{min} = 5.87(7)  \;\;,\;\;\;  \theta_{(+,+)}(x_{min}) =  -1.75(1)   \;.
\end{equation}
Also these estimates are fully consistent with those of our previous work 
\cite{MHcorrections}.

\section{Thermodynamic Casimir force for $(O,+)$ boundary conditions}
\label{mixedresults}
The three-dimensional Ising model and the improved Blume-Capel model 
with $(O,+)$ boundary conditions have been simulated in refs. 
\cite{VaGaMaDi08}   and \cite{mycrossing,PaTrDi13}, respectively.
In ref. \cite{mycrossing} we simulated films with $(0,+)$ boundary conditions 
for the thicknesses  $L_0=8.5$, $12.5$, and $16.5$ by using a combination of heat-bath and cluster
updates. As transversal extension we took $L=32$, $48$, and $64$, respectively. Note that the correlation 
length of the film is $\xi_{2nd,Film} \approx 0.224 (L_0+ L_s) $ at the critical point \cite{mycrossing}. 
Therefore we expect that finite $L$ effects are small for the values that we had chosen.
We performed $10^8$, $10^8$, and $2 \times 10^8$ update cycles for $L_0=8.5$, $12.5$, and $16.5$,
respectively. In total 10 years of CPU time on a single core of an AMD Opteron 2378 were used.

Here we complement these simulations and study the thicknesses $L_0=16.5$ and $L_0=24.5$ 
using $L=64$ and $96$, 
respectively.  We used the same type of update-cycle as above for $(+,+)$ and $(+,-)$ boundary 
conditions. In particular we used $i_r=2$ and $n_{exc} = 20$ for $L_0= 16.5$ and
$i_r=3$ and $n_{exc} = 20$ for $L_0= 24.5$. For each value of $\beta$ we simulated at, 
$10^7$ update cycles were performed.  This large number of updates, compared with the study 
of $(+,-)$ and $(+,+)$ boundary conditions discussed above, is needed to get accurate
results for the first and second derivative of the thermodynamic Casimir force with respect
to the surface field $h_1$.  Also these simulations took about 10 years of CPU time on a single core 
of an AMD Opteron 2378.

\begin{figure}
\begin{center}
\includegraphics[width=14.5cm]{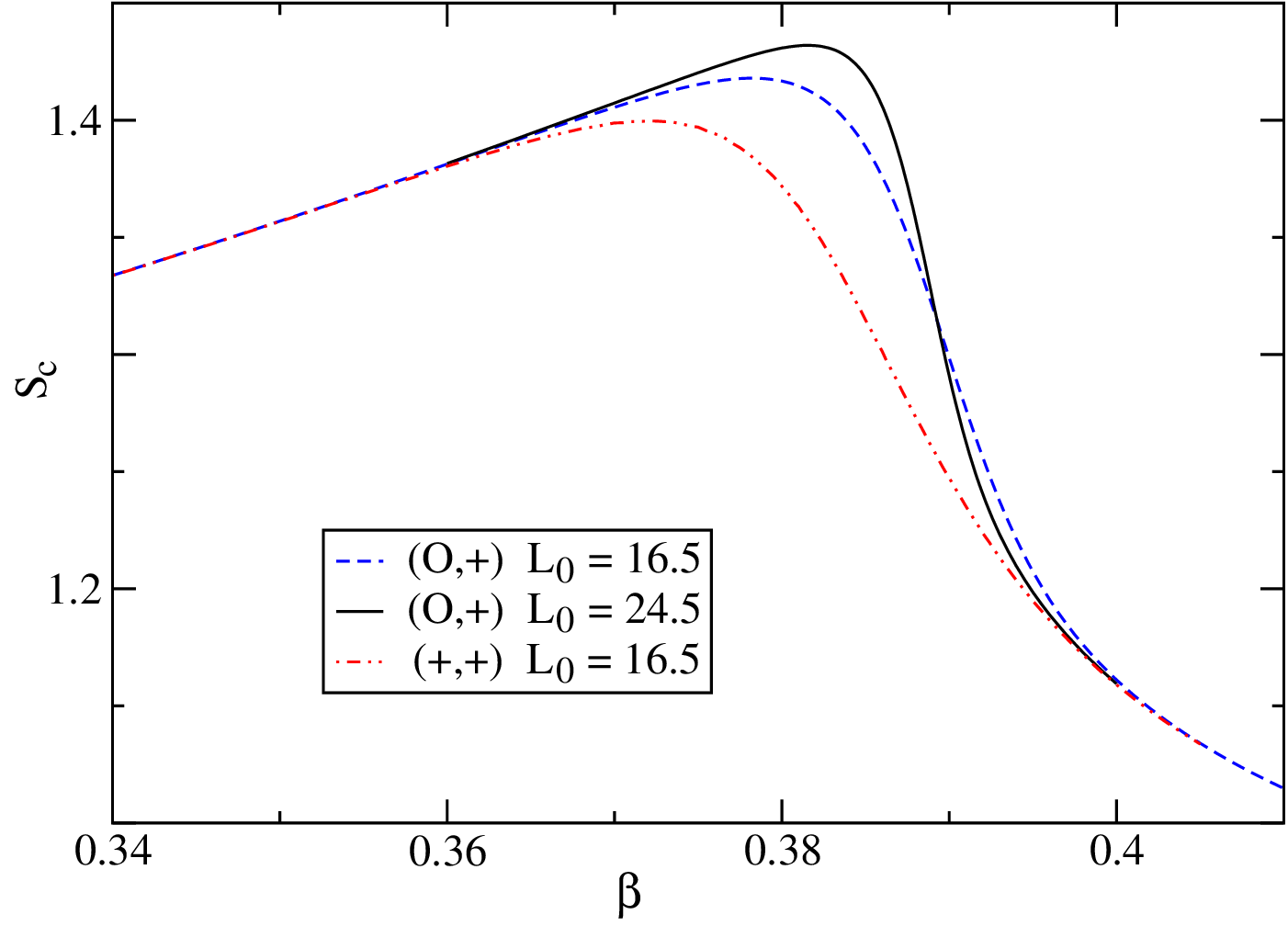}
\caption{\label{ballmix}
We plot the average size $S_c$ per area of the frozen exchange clusters  as a function of $\beta$
for $(O,+)$ boundary conditions and the thicknesses $L_0=16.5$ and $32.5$ of the film.
For comparison we give  $S_c$ for $(+,+)$ boundary conditions and $L_0=16.5$.
}
\end{center}
\end{figure}

In the case of $(O,+)$ boundary conditions we have the choice, whether we perform the 
exchange cluster update at the $+$ or the $O$ boundary. Taking the conventions of sections
\ref{filmgeometry} and \ref{algorithm}, this means that we either fix $s_{x,1} = s_{x,2}=0$ for $x_0=0$,
$s_{x,1} = 1$ for $x_0=L_0+3/2$ and $s_{x,2} = 1$ for $x_0=L_0+1/2$  or 
$s_{x,1} = s_{x,2}=1$ for $x_0=0$,
$s_{x,1} = 0$ for $x_0=L_0+3/2$ and $s_{x,2} = 0$ for $x_0=L_0+1/2$. In both cases, 
the frozen clusters have their origin at $x_0=L_0+1/2$. 
Preliminary tests show  that it is preferential to perform the exchange cluster algorithm 
at the $+$ boundary. 
In Fig. \ref{ballmix} we give the  average size $S_c$ per area of the frozen exchange clusters
for $(O,+)$ boundary conditions, where the exchange cluster update is performed at
the $+$ boundary.  For comparison we give the analogous result for 
$(+,+)$ boundary conditions and $L_0=16.5$. At high and low values of $\beta$,
$S_c$ does not depend on the thickness of the film.  Furthermore it coincides 
with $S_c$ for $(+,+)$ boundary conditions. In the neighborhood of $\beta_c$ the 
behavior of $S_c$ depends on $L_0$ and furthermore for $L_0=16.5$, the behavior 
for $(+,+)$ and $(O,+)$ boundary conditions is different.
We notice that also for $(O,+)$ boundary conditions, $S_c$ remains small compared
with the thickness $L_0$ of the film in the whole range of $\beta$ that we have 
simulated.   

For comparison, we simulated for $L_0=16.5$
with the exchange cluster update performed at the $O$ boundary at 41 values 
of $\beta$, and $4 \times 10^5$ update cycles only. In Fig. \ref{ballOO16}  we  plot the resulting $S_c$.
 We see that $S_c$ assumes a maximum $\approx 2.12$ at $\beta \approx \beta_c$, 
which is considerably larger than the maximum $\approx 1.42$
for the other choice, reached at $\beta \approx 0.38$. At $\beta=0.34$, which is the smallest 
inverse temperature that we simulated, $S_c$ is almost equal for the two choices.
On the other hand 
for $\beta=0.41$, $S_c \approx 1.51$ for the exchange cluster performed at the $O$ boundary, 
while $S_c \approx 1.11$ for the exchange cluster performed at the $+$ boundary. 

\begin{figure}
\begin{center}
\includegraphics[width=14.5cm]{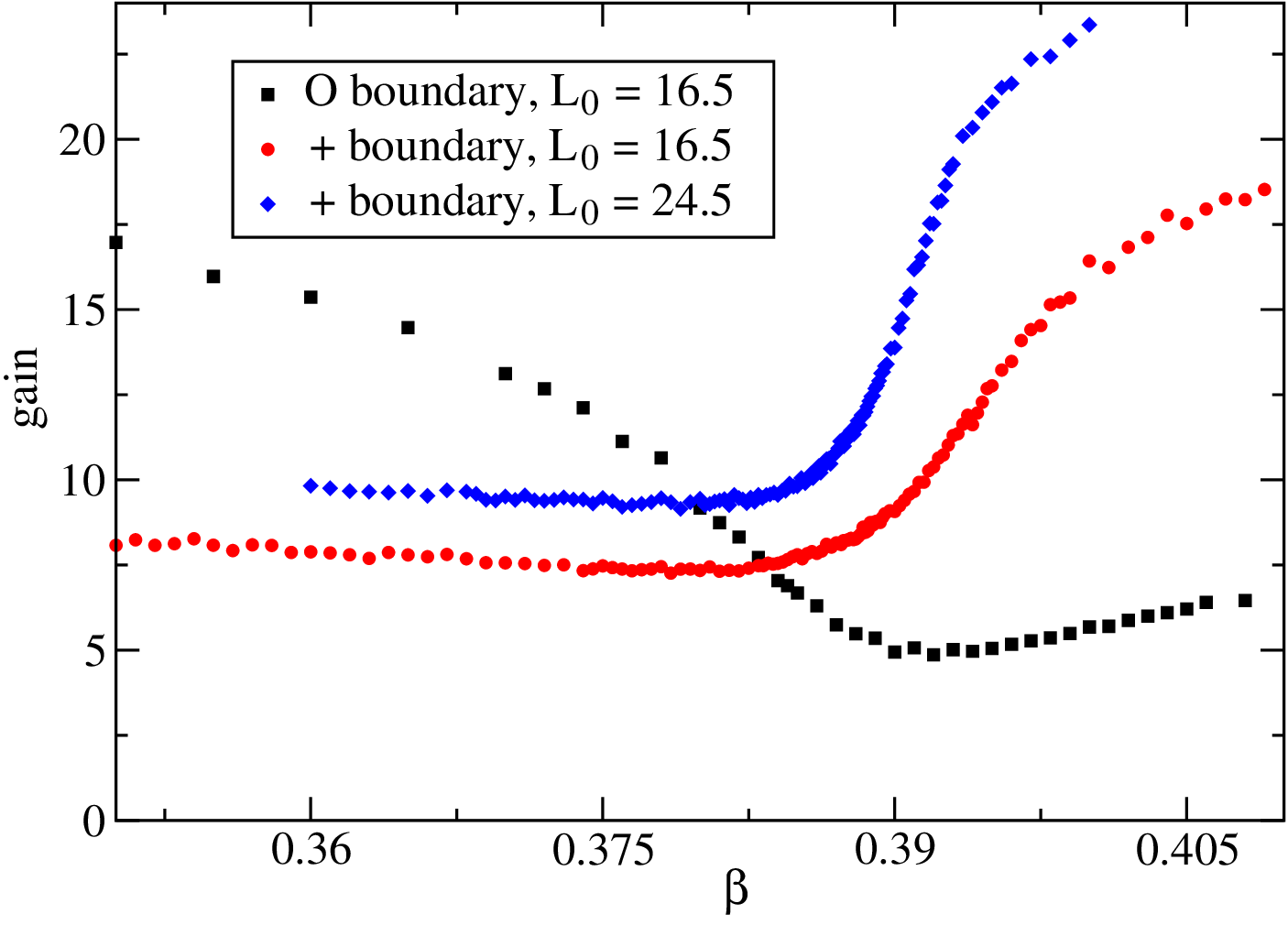}
\caption{\label{gainmix}
We plot the gain for films with $(O,+)$ boundary conditions. For $L_0=16.5$ we performed 
the exchange cluster update at the $O$ as well as the $+$ boundary. For  $L_0=24.5$
only exchange cluster updates at the $+$ boundary were performed.
}
\end{center}
\end{figure}

In Fig. \ref{gainmix} we plot gain~(\ref{gaindef}) as a function of $\beta$. For $L_0=16.5$ 
we give results for both performing the exchange cluster update at the $O$ as well as the $+$ boundary.
For $L_0=24.5$ only results for  performing the exchange cluster update at the $+$ boundary
are available.  The behavior of gain for the exchange cluster updates at the $+$ boundary is 
qualitatively very similar to what we have seen above for $(+,+)$ boundary conditions. For 
$\beta \lessapprox \beta_c$ it depends little on $\beta$, while for larger values of $\beta$ we see 
a rapid increase of the gain with increasing $\beta$.  The behavior for the exchange cluster updates at the $O$ 
boundary is complementary. For $\beta \lessapprox \beta_c$, the gain increases with decreasing 
$\beta$, while for larger values of $\beta$ we see only a small increase with increasing $\beta$.
The intersection between the two gain curves for $L_0=16.5$ is located at $\beta \approx 0.383$, where 
$\xi = 6.643(1)$ \cite{MHamplitude}. Overall, also taking into account the behavior of $S_c$, 
performing the cluster exchange algorithm at the $+$ boundary is the better choice. Both
versions of the cluster update clearly reduce the variance of $\Delta E$.

\begin{figure}
\begin{center}
\includegraphics[width=14.5cm]{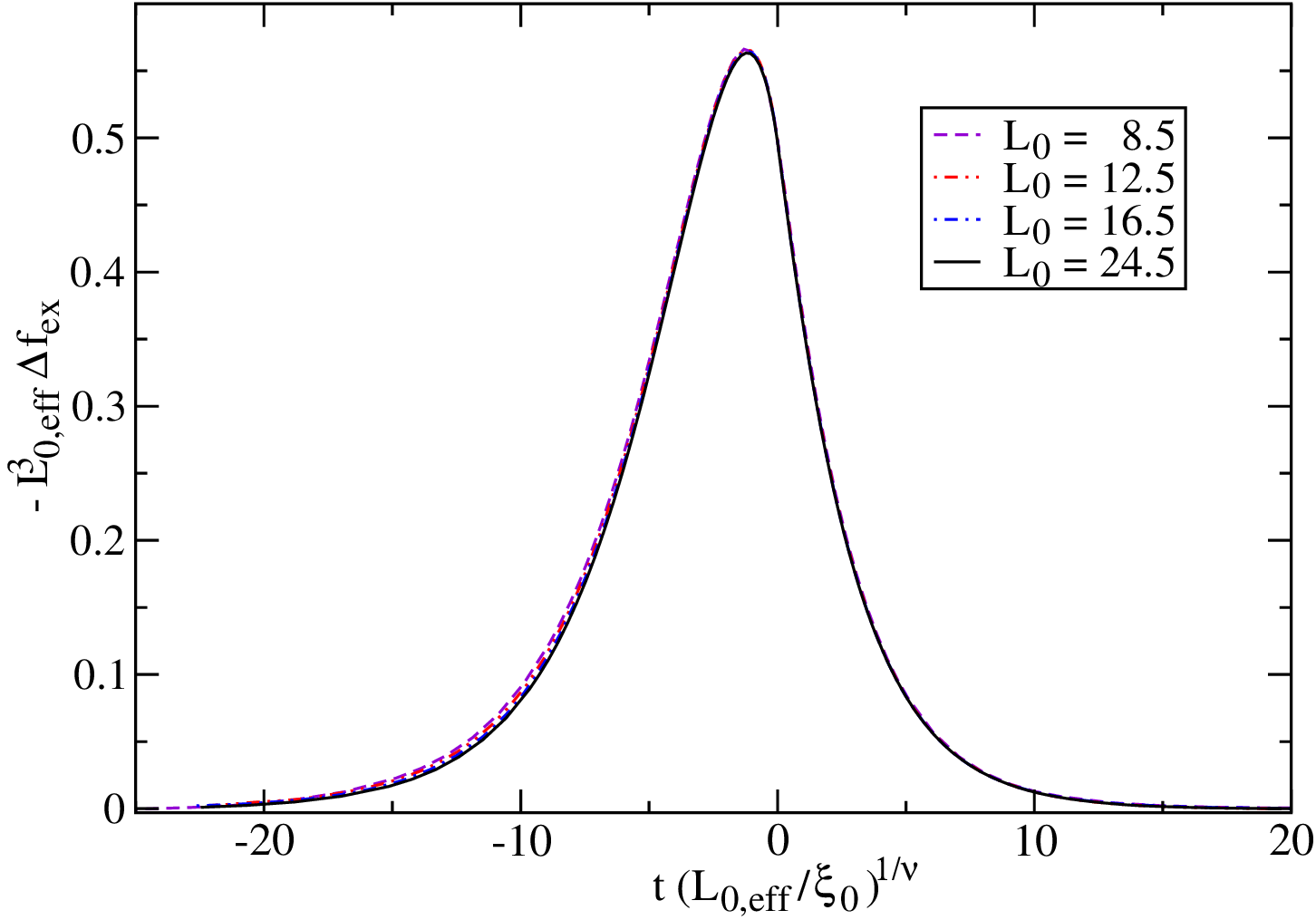}
\caption{\label{mixplot}
Numerical results for the scaling function $\theta_{(O,+)}(x)$.
We plot $-L_{0,eff}^3 \Delta f_{ex}$ as a function of $t [L_{0,eff}/\xi_0]^{1/\nu}$, where
$L_{0,eff} = L_0 +L_s$ with $L_s=1.43$, $\xi_0=0.2283$, and $\nu=0.63002$. The thicknesses
of the films are $L_0 = 8.5$, $12.5$, $16.5$, and $24.5$. The error bars are typically smaller than
the thickness of the lines.
}
\end{center}
\end{figure}

Let us discuss the results for the scaling function of the thermodynamic Casimir force.
In Fig. \ref{mixplot} we plot our numerical results for $\theta_{(O,+)}(x)$. The data for 
$L_0=8.5$ and $12.5$ are taken from ref. \cite{mycrossing}, while those for $L_0=16.5$ and $24.5$ 
are computed by using the exchange cluster algorithm. For $x \gtrapprox -5$ the curves fall perfectly
on top of each other. For smaller values of $x$, small differences between the results for different 
thicknesses can be observed. The scaling function function $\theta_{(O,+)}(x)$
shows a maximum in the low temperature phase, very close to the critical point. In order to locate
the maximum, we determine the zero of $\Delta E_{ex}$. We find $\beta_{max}$=$0.390713(6)$, $0.389446(6)$,  
$0.3888747(15)$ and $0.3883626(10)$, for $L_0=8.5$, $12.5$, $16.5$ and $24.5$, respectively.
This corresponds to $x_{max}=t_{max} [(L_0 + L_s)/\xi_0]^{1/\nu} = $
$-1.1925(24)[38]$, $-1.1764(41)[27]$, $-1.1743(15)[21]$, and $-1.1723(18)[14]$.   
For  $\theta_{(O,+)}(x_{max})$  we get the estimates $- \Delta f_{ex}(\beta_{max}) [L_0+L_s]^3 =$
$0.5664(7)[34]$, $0.5657(5)[24]$, $0.5647(4)[19]$, and $0.5635(4)[13]$, where we used $L_s=1.43(2)$ as 
input. The number in $[]$ gives the error due to the uncertainty of $L_s$.
We see that $- \Delta f_{ex}(\beta_{max}) [L_0+L_s]^3$  is  monotonically decreasing with $L_0$ and
the error due to the uncertainty of $L_s$ is larger than the statistical one. Therefore we performed
a fit, leaving $L_s$ as free parameter. We get, taking all four thicknesses into account, 
$\theta_{(O,+)}(x_{max}) =0.5636(23)$  and $L_s =1.41(2)$, which is consistent with our previous estimate of 
$L_s$.  As our final estimate we quote
\begin{equation}
 x_{max} = -1.168(5) \;\;\;, \;\;\;   \theta_{(O,+)}(x_{max}) = 0.5635(20) \;\;
\end{equation}
where we extrapolated $x_{max}$ linearly in $L_0^{-2}$ to $L_0 \rightarrow \infty$.
The error bar of $x_{max}$ is chosen such that the estimate obtained for $L_0=24.5$ is included.
In the case of $\theta_{(O,+),max}$ the estimate obtained for $L_0=24.5$ and our fit essentially 
coincide, which leads to our final estimate. Our present estimates are compatible with
$x_{max} = -1.174(10)$ and $\theta_{(O,+),max} = 0.564(3)$, ref. \cite{mycrossing}, and the error bars
are slightly reduced. 
For a summary of previous results we refer the reader to section VI C of
ref. \cite{mycrossing}.  At the critical point we get $- \Delta f_{ex}(\beta_{max}) [L_0+L_s]^3 =$
$0.4978(7)[30]$, $0.4982(6)[21]$, $0.4976(4)[17]$, and $0.4964(3)[11]$ for $L_0=8.5$, $12.5$, 
$16.5$ and $24.5$, respectively, where again we used $L_s=1.43(2)$ as input. The value for $L_0=24.5$
is slightly smaller than that for $L_0=8.5$, $12.5$, and $16.5$.  Mainly based on the 
result for $L_0=24.5$ we quote 
\begin{equation}
\theta_{(O,+)}(0) = 0.496(2)
\end{equation}
as our final result, which is fully consistent with $\theta_{(O,+)}(0) = 0.497(3)$ obtained in 
ref. \cite{mycrossing} and with 
$\theta_{(O,+)}(0) = 0.492(5)$ given in eq.~(34) of ref. \cite{PaTrDi13}.

Next let us turn to the derivatives of the thermodynamic Casimir force per area
with respect to the surface field $h_1$. The thermodynamic Casimir force per area
as a function of the inverse temperature $\beta$ and the surface field $h_1$
follows the scaling law
\begin{equation}
\label{generalizedFSS}
F_{Casimir}(\beta,h_1) =  k_B T L_0^{-d} \Theta_{(O,+)}(x, x_{h_1})
\end{equation}
where 
\begin{equation}
\label{defxh}
 x_{h_1} = h_1 [L_0/l_{ex,nor,0}]^{y_{h_1}} \;
\end{equation}
where for our model $l_{ex,nor,0}=0.213(3)$, eq. (73) of \cite{mycrossing}, and the 
surface critical RG-exponent $y_{h_1} = 0.7249(6)$ , eq. (52) of \cite{mycrossing}.
In particular for a vanishing surface field we get the scaling function
\begin{equation}
\theta_{(O,+)}(x) = \Theta_{(O,+)}(x,0) \;\;
\end{equation}
discussed above.

Following ref. \cite{mycrossing}, 
we compute the Taylor-expansion  of the thermodynamic Casimir
force with respect to the boundary field $h_1$ around $h_1=0$ up to the second
order. To this end we compute the first and second derivative of
$\Delta f_{ex}$ with respect to $h_1$. The $n^{th}$ derivatives can be
written as
\begin{equation}
\frac{\partial^{n} \Delta f_{ex}(L_0,\beta,h_1)}{\partial h_1^{n}} =
-\int_{\beta_0}^{\beta} \mbox{d} \tilde \beta
 \frac{\partial^{n} \Delta E_{ex}(L_0, \tilde \beta,h_1)}{\partial h_1^{n}}
\end{equation}
where
\begin{equation}
  \frac{\partial^n \Delta E_{ex}(L_0, \beta,h_1)}{\partial h_1^{n}}
= \frac{\partial^n \langle E \rangle_{(L_0+1/2, \beta,h_1)} \rangle}{\partial h_1^n}
- \frac{\partial^n \langle E \rangle_{(L_0-1/2, \beta,h_1)} }{\partial h_1^n} \;\;.
\end{equation}
Note that there is no bulk contribution, since the internal energy of the bulk
does not depend on $h_1$. In the Monte Carlo simulation, the first derivative
can be computed as
\begin{equation}
\label{firstderiv}
\frac{ \partial \langle E \rangle_{(L_0, \beta,h_1)}}{\partial h_1}
= \langle E M_1 \rangle - \langle  E \rangle \langle M_1 \rangle
\end{equation}
where
\begin{equation}
 M_1 =  \sum_{x_1,x_2} s_{(1,x_1,x_2)}  \;\;.
\end{equation}
The second derivative is given by
\begin{equation}
\label{secondderiv}
\frac{\partial^2 \langle E \rangle_{(L_0, \beta,h_1)}}{\partial h_1^2}
 = \langle E M_1^2 \rangle
 - 2 \langle  E  M_1 \rangle \langle M_1 \rangle
  - \langle  E  \rangle \langle M_1^2 \rangle
+ 2 \langle  E \rangle \langle M_1 \rangle^2 \;\;.
\end{equation}
Higher derivatives could be computed in a similar way. However it turns
out that the relative statistical error of the second derivative is much larger
than that of the first one. Therefore we abstain from implementing 
higher derivatives.

We computed the quantities~(\ref{firstderiv}, \ref{secondderiv})
with reduced variance by using the exchange cluster update. Here 
we did not work out an explicit expression as eq.~(\ref{improvedE}) for $\Delta E$.
Instead we implemented eq.~(\ref{improved_general}) directly for the observables
that enter eqs.~(\ref{firstderiv},\ref{secondderiv}). In order to avoid a numerical
effort that is proportional to the volume of the film, we kept track
of the values of $E$ of the two films, while exchange cluster updating 
and performing the Todo-Suwa updates of the $i_r$ layers. 

In the case of $L_0=16.5$ we can compare with our results obtained in 
ref. \cite{mycrossing}, where we performed 20 times more measurements.
Using the cluster exchange update, we have reduced the statistical error of 
$\frac{\partial \Delta E_{ex}(L_0, \tilde \beta,h_1)}{\partial h_1}$
by a factor slightly larger than $2$ for $\beta \lessapprox \beta_c$
compared with the result of ref. \cite{mycrossing}. In the 
low temperature phase this factor increases up to $\approx 6$ at 
$\beta=0.405$. For the second derivative with respect to $h_1$ a similar
reduction of the statistical error can be observed.

\begin{figure}
\begin{center}
\includegraphics[width=14.5cm]{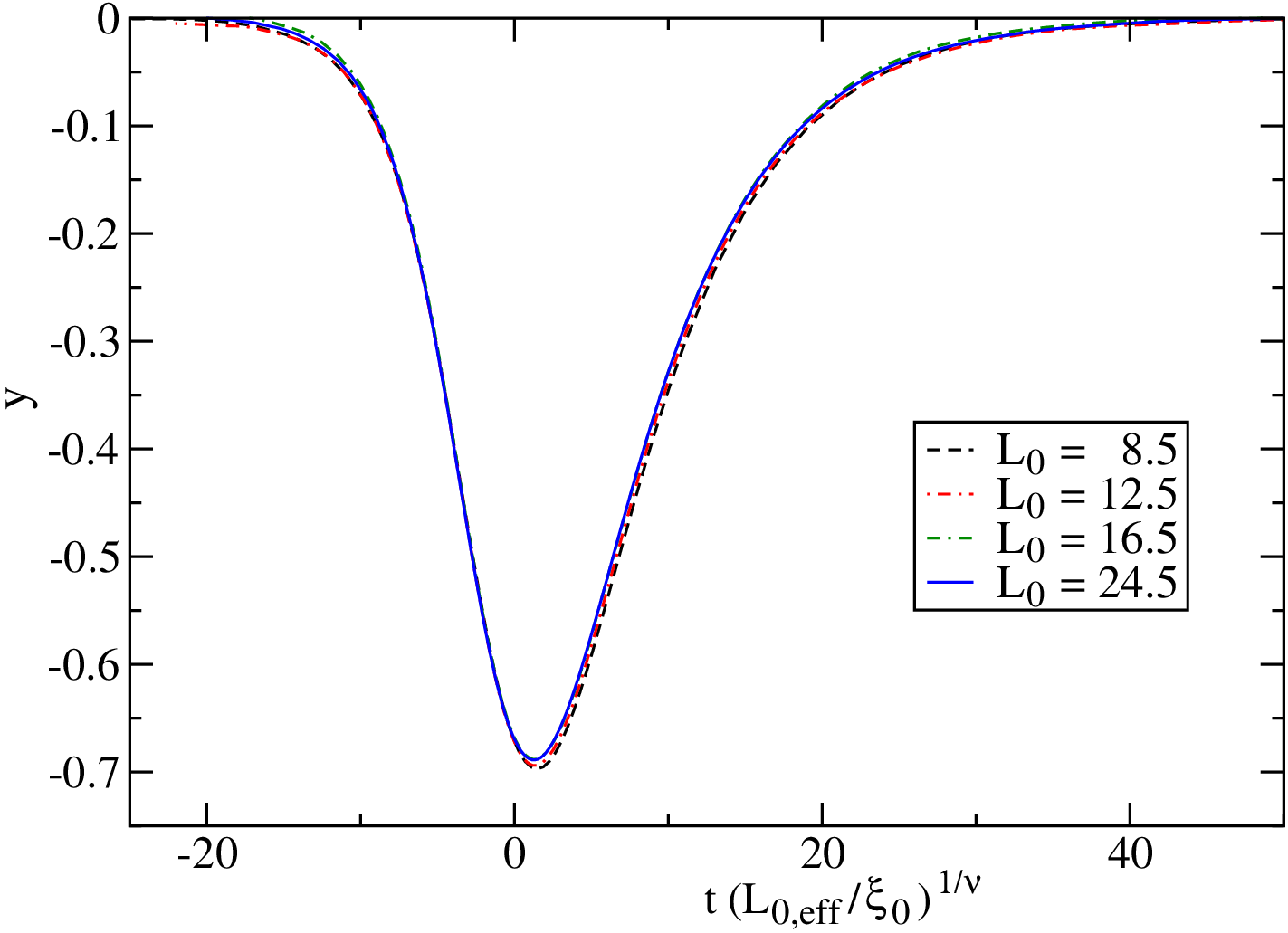}
\caption{\label{mixploth} We plot
$y=  - L_{0,eff}^3 (L_{0,eff}/l_{ex,nor,0})^{-y_{h_1}}
 \frac{\partial \Delta f_{ex}}{\partial h_1}$ as
a function of $ t (L_{0,eff}/\xi_0)^{1/\nu}$ for $(0,+)$ boundary conditions
for the thicknesses $L_0=8.5$, $12.5$, $16.5$, and $24.5$. To this end,
we have used $L_{0,eff}=L_0+L_s$ with $L_s=1.43$, $\xi_0=0.2283$,
$\nu = 0.63002$, $l_{ex,nor,0}=0.213$, and $y_{h_1}=0.7249$.
}
\end{center}
\end{figure}

In Fig. \ref{mixploth} we plot our results for 
$\theta_{(O,+)}'(x) \equiv \left . \frac{\partial \Theta_{(O,+)}(x,x_{h_1})}{\partial h_1}
\right |_{h_1=0}$. The curves for 
different thicknesses fall nicely on top of each other. One observes that in 
contrast to $\theta(x)$, $\theta'(x)$ has a large amplitude also for
$x \ge 0$. In particular the minimum is located close
to the critical point, in the high temperature phase.
The analysis of the data gives
$\beta_{min} = 0.38404(5)$, $0.38575(3)$, $0.38644(2)$ and $0.387020(10)$ for $L_0=8.5$, $12.5$, 
$16.5$ and $24.5$, respectively.  This corresponds to 
$x_{min} =$ $1.468(20)[5]$, $1.345(20)[3]$,  $1.305(20)[2]$, and  $1.284(18)[2]$, where again the 
number in $[]$ gives the error due to the uncertainty of $L_s$. Still we see a small trend in the 
numbers. Therefore we extrapolated linearly in $1/L_0^2$, arriving at $x_{min} =1.253(16)$.
For $- L_{0,eff}^3 (L_{0,eff}/l_{ex,nor,0})^{-y_{h_1}}
 \frac{\partial \Delta f_{ex}}{\partial h_1}$, with $L_{0,eff}=L_0+L_s$ we get at the minimum
the values $-0.697(1)[3]\{7\}$, $-0.694(2)[2]\{7\}$, $-0.691(1)[2]\{7\}$, and $-0.689(1)[1]\{7\}$  for 
$L_0=8.5$, $12.5$, $16.5$ and $24.5$, respectively.  Here the number in $[]$, gives again the error
due to the uncertainty of $L_s$, while the number in $\{\}$ gives the error induced by the 
uncertainty of $l_{ex,nor,0}$. It turns out that the latter is dominating.  As our final result
we quote
\begin{equation}
 x_{min} = 1.25(4) \;\;\;, \;\;\;   \theta_{(O,+)}'(x_{min}) = -0.689(3)\{7\} \;\;.
\end{equation}
As final estimate of $x_{min}$ we took our extrapolation and the error bar is chosen such that 
the result for $L_0=24.5$ is still included. As final estimate of $\theta_{(O,+),min}'$ we 
simply took the result obtained for $L_0=24.5$. The error bar given in $()$ 
is mainly motivated by the comparison with the result for $L_0=16.5$. 
The dominant error given in $\{\}$ is due to the uncertainty of $l_{ex,nor,0}$.
Our present estimates are consistent with and slightly more accurate than those 
given in ref. \cite{mycrossing}.

\begin{figure}
\begin{center}
\includegraphics[width=14.5cm]{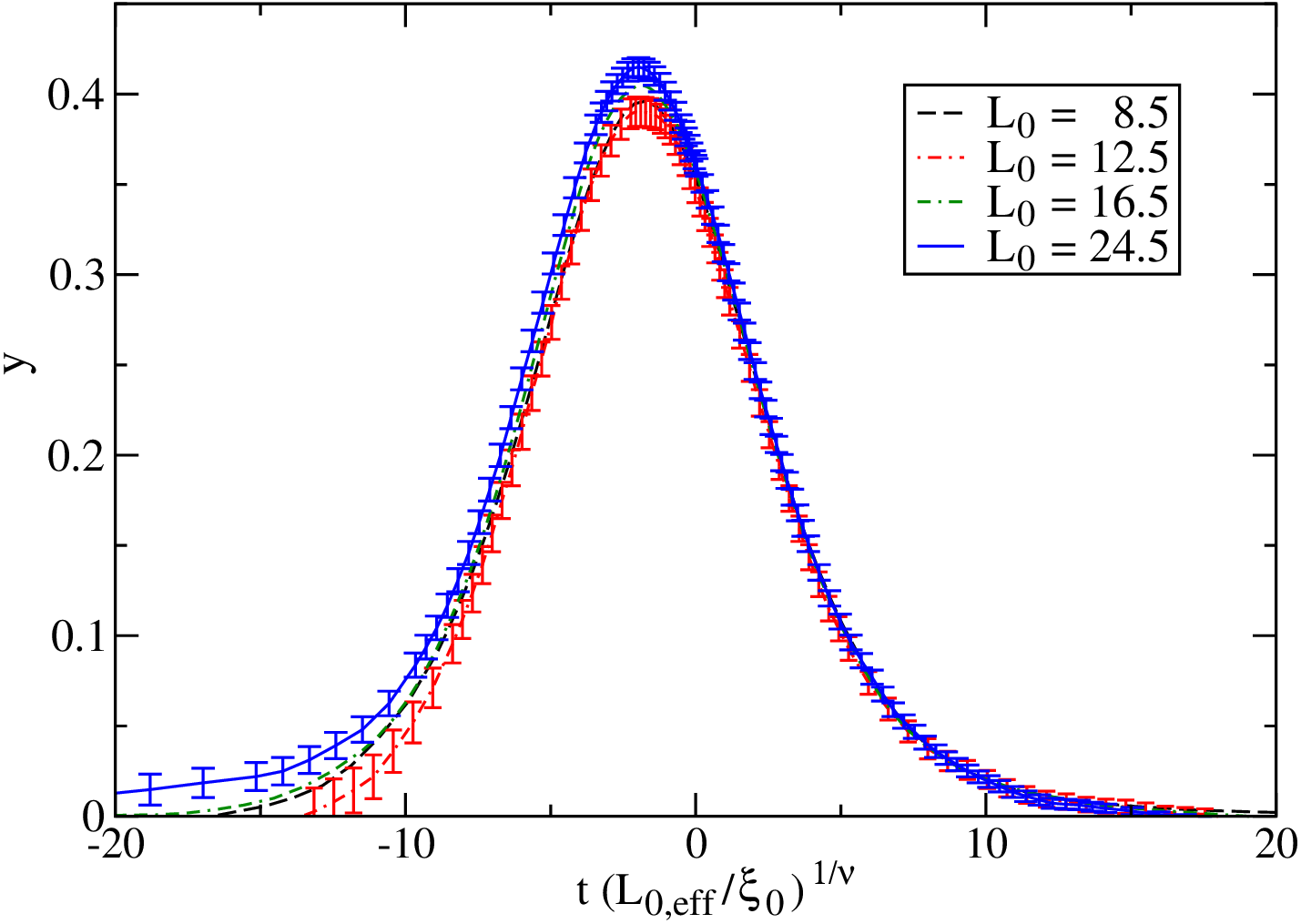}
\caption{\label{mixplothh} We plot
$y=  - L_{0,eff}^3 (L_{0,eff}/l_{ex,nor,0})^{-2 y_{h_1}}
 \frac{\partial^2 \Delta f_{ex}}{\partial h_1^2}$ as
a function of $ t (L_{0,eff}/\xi_0)^{1/\nu}$ for $(0,+)$ boundary conditions
for the thicknesses $L_0=8.5$, $12.5$, $16.5$, and $24.5$. To this end,
we have used $L_{0,eff}=L_0+L_s$ with $L_s=1.43$, $\xi_0=0.2283$,
$\nu=0.63002$, $l_{ex,nor,0}=0.213$, and $y_{h_1}=0.7249$.
}
\end{center}
\end{figure}

In Fig. \ref{mixplothh} we plot our results for 
$\theta_{(O,+)}''(x) \equiv \left . \frac{\partial^2 \Theta_{(O,+)}(x,x_{h_1})}{\partial h_1^2}
\right |_{h_1=0}$. Here the error bars 
are, despite of the variance reduction, larger than the thickness of the lines.
For $L_0=12.5$, taken from ref. \cite{mycrossing}, and $L_0=24.5$ we give the 
error bars. For $L_0=8.5$ and $16.5$ we omit them to keep the figure readable.
The curves for different thicknesses fall reasonably well on top of each other.
The discrepancies might be attributed to the statistical error.  
The function displays a single maximum.  Analysing the data we arrive 
at the final result
\begin{equation}
 x_{max} = -2.0(1) \;\;\;, \;\;\;   \theta_{(O,+)}''(x_{max}) = 0.41(1)\{1\} \;\;.
\end{equation}
The number given in $\{\}$
gives the error due to the uncertainty of $l_{ex,nor,0}$. Again 
our result is consistent with ref. \cite{mycrossing}.

We have demonstrated that also the statistical error of the derivatives
of $\langle \Delta E \rangle$ with respect to the boundary field $h_1$ can be reduced 
by using the exchange cluster update. As a result, we reduced the errors
of the scaling function $\theta_{(O,+)}(x)$, $\theta_{(O,+)}'(x)$, 
and $\theta_{(O,+)}''(x)$ with respect to ref. \cite{mycrossing}.  This 
however leaves the conclusions of ref. \cite{mycrossing} unchanged.
Therefore we refer the reader to ref. \cite{mycrossing} for a detailed
discussion. A particularly interesting 
observation is that for a finite boundary field $h_1$ the thermodynamic
Casimir force might change sign as a function of the thickness $L_0$.
\section{Films with $(O,O)$ boundary conditions}
In contrast to the  cases studied above, $(O,O)$ boundary conditions
do not break the global $\mathbb{Z}_2$ symmetry of the system. 
Therefore films with $(O,O)$ boundary conditions 
are expected to undergo a second order phase transition that belongs
to the universality class of the two-dimensional Ising model.
At this transition the correlation length of the film diverges and we therefore expect
large finite size effects, where the finiteness in the transversal directions is meant. 
This should also effect the thermodynamic Casimir force. This problem has been
discussed in ref. \cite{VaGaMaDi08} and for the case of films with periodic boundary conditions 
in ref. \cite{HuGrSc11}. Here we put this discussion on a quantitative level.
Since the transition belongs to the two-dimensional Ising universality
class, we can make use of the universal finite size scaling function
of the free energy density that we compute below in  section \ref{finite2D} by using the
exact solution of the two-dimensional Ising model \cite{Kaufman}. In section \ref{filmtc},
in order to  make use of this universal function, we accurately determine the
transition temperature and match the scaling variable for a large range of thicknesses
of the film.  Finally in section \ref{casimiroo} we compute the thermodynamic 
Casimir force for $L_0=8.5$, $12.5$, $16.5$, and $24.5$ by using the exchange cluster
algorithm. The algorithm seems to fail in reducing the variance in the low 
temperature phase of the films. We suggest to remediate this problem by breaking 
by hand the $\mathbb{Z}_2$ symmetry in the low temperature phase. Still, in the
neighborhood of the transition of the film, we benefit only little from the 
exchange cluster update.

\subsection{Finite size effects in the neighborhood of the 2D transition}
\label{finite2D}
The reduced Hamiltonian of the Ising model on the square lattice 
in the absence of an external field is given by
\begin{equation}
 H = - \beta \sum_{<xy>}   s_x s_y
\end{equation} 
where $s_x \in \{-1,1\}$ and $<xy>$ is a pair of nearest neighbor sites.
For the discussion of the critical behavior of the Ising model on the square lattice it
is convenient to introduce
\begin{equation}
\label{reduced2D}
 \tau = \frac{1}{2} \left( \frac{1}{\sinh 2\beta} - \sinh 2\beta \right)
\end{equation}
as reduced temperature.
The exponential correlation length in the thermodynamic limit behaves as
\begin{equation}
\label{xi2DIsing}
 \xi \simeq \xi_{0,\pm} |\tau|^{-\nu}
\end{equation}
where $\nu=1$, $\xi_{0,+}=1/\sqrt{2}$ and $\xi_{0,-}=\xi_{0,+}/2$,
where $\xi_{0,+}$ and $\xi_{0,-}$ are the amplitudes of the
exponential correlation length in the high and the low temperature
phase, respectively.

The reduced free energy density in the thermodynamic limit
is given by  \cite{McCoyWu}  
\begin{equation}
\label{freeLimit}
f(\tau) = - \frac{1}{2} \ln (2 \cosh^2 2 \beta)  + f_{sing} (\tau)
\end{equation}
where
\begin{equation}
 f_{sing} (\tau) = - \int_{0}^{\pi} \frac{\mbox{d} \theta}{2 \pi} \;
 \ln \left[ 1 + \left(1 -  \frac{\cos^2 \theta }{1 + \tau^2} \right)^{1/2} \right] \; .
\end{equation}
In the neighborhood of the critical point, the reduced free energy density behaves as
\begin{equation}
 f(\tau) \simeq \frac{1}{2 \pi} \tau^2 \ln |\tau| + A(\tau)
\end{equation}
where $A(\tau)$ is an analytic function.

\begin{figure}
\begin{center}
\includegraphics[width=15cm]{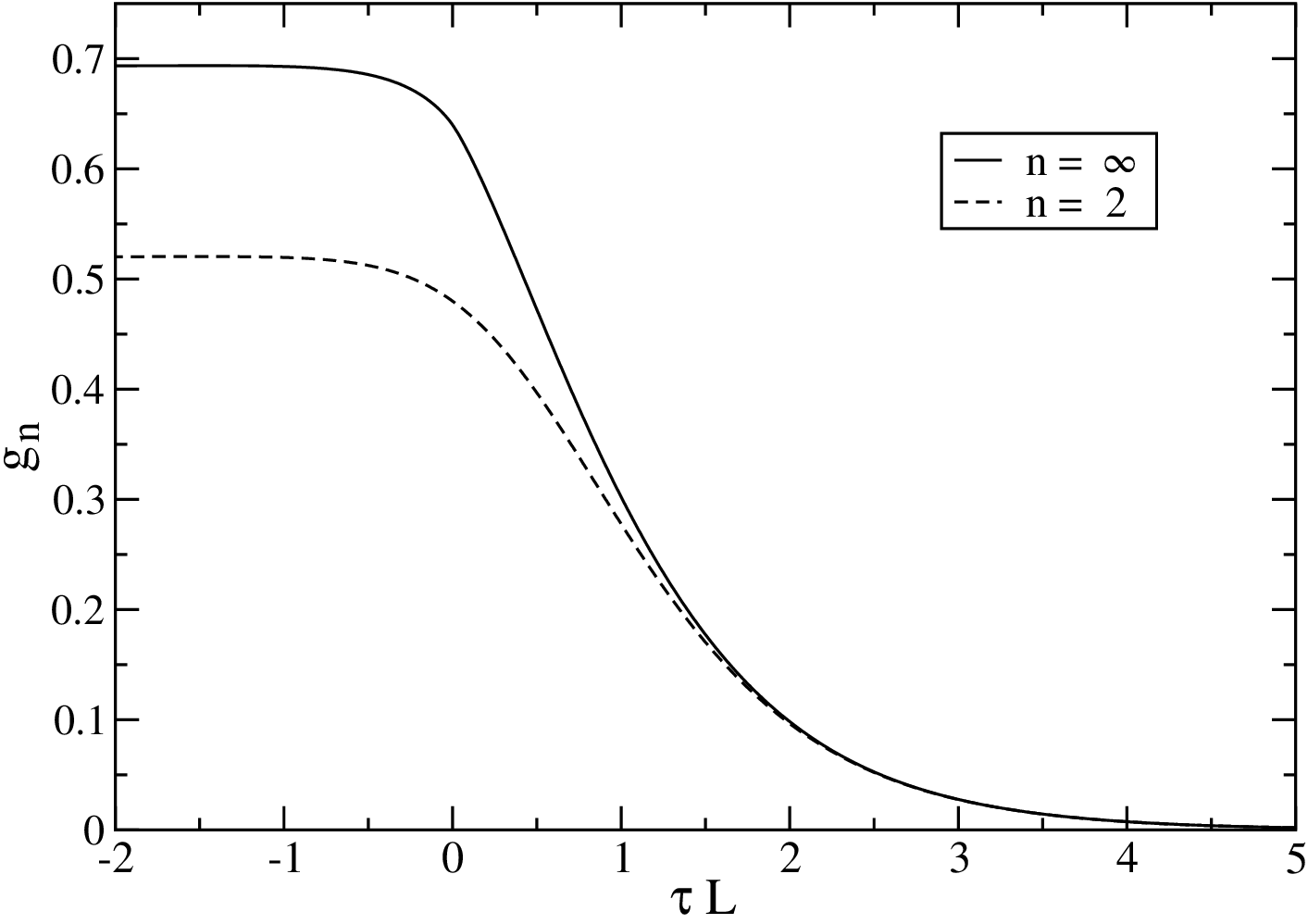}
\caption{\label{freeplot}
Finite size scaling function $g_n(\tau L)$ obtained by evaluating 
the free energy density of the Ising model on the square lattice with 
periodic boundary conditions for $L=1024$. 
}
\end{center}
\end{figure}

Here we are interested in the finite size scaling behavior of the reduced 
free energy density
\begin{equation}
 f(\beta,L) = - \frac{1}{L^2} \ln Z(\beta,L)
\end{equation}
where $L=L_1=L_2$ is the linear extension of the lattice and periodic
boundary conditions are assumed. To this end
we have numerically evaluated  eq.~(39) of ref. \cite{Kaufman}.
The differences 
\begin{equation}
 \Delta f_2(\beta,L) =  f(\beta,2 L) - f(\beta,L)
\end{equation}
and 
\begin{equation}
 \Delta f_{\infty}(\beta,L) =  f(\beta,\infty) - f(\beta,L)
\end{equation}
are governed by  finite size scaling functions
\begin{equation}
\label{defgn}
 g_n(\tau L) \simeq \Delta f_n(\beta,L)  L^{2}  \;.
\end{equation}
We have constructed the function $g_2$  numerically by 
evaluating eq.~(39) of ref. \cite{Kaufman}. In order to get
$g_{\infty}$, eq.~(\ref{freeLimit}) is used in addition.
Our results obtained for $L=1024$ are given in fig. \ref{freeplot}.
Comparing with results for smaller $L$, we conclude that the 
deviation of our result for $L=1024$ from the asymptotic limit is 
less than $10^{-6}$.
\subsection{The phase transition of films with $(O,O)$ boundary conditions}
\label{filmtc}
The transition is expected to be of second order and to share the universality class 
of the two-dimensional Ising model. This allows us to take advantage of 
exact results obtained for the two-dimensional Ising and conformal 
field theory. In our numerical study we shall 
follow the approach of ref. \cite{CaHa96}, where films of the Ising 
model with periodic boundary conditions were studied.

We determine the inverse transition temperature $\beta_{c,2D}(L_0)$ by 
finite size scaling. For simplicity we consider lattices with $L_1=L_2=L$. 
An estimate $\bar{\beta}_{c,2D}(L_0,L)$ of $\beta_{c,2D}(L_0)$ is given
by the solution of
\begin{equation}
\label{determinebc}
  R(\beta,L_0,L) = R^*
\end{equation}
where $R(\beta,L_0,L)$ is a renormalization group invariant quantity like 
the Binder cumulant $U_4$, the second moment correlation length over 
the lattice size $\xi_{2nd}/L$ or the ratio of partition functions $R_Z=Z_a/Z_p$,
where $Z_a$ is the partition function of a system with periodic boundary
conditions in 1-direction and anti-periodic boundary conditions in 2-direction,
while $Z_p$ is the partition function of a system with periodic boundary in 
both 1 and 2-direction. The fixed point value $R^*$ is defined by
\begin{equation}
 R^* := \lim_{L \rightarrow \infty}  R(\beta_{c,2D},L_0,L) \;\;.
\end{equation}
It can be obtained, e.g. from the study of the two-dimensional Ising model.
It is known to high numerical precision for $\xi_{2nd}/L$ and $U_4$ \cite{SaSo00}.
The fixed point value of $R_Z$ is  exactly
known for arbitrary ratios $L_1/L_2$. It can be derived both from the exact 
solution of the two-dimensional Ising model \cite{Kaufman} as well as from conformal field 
theory. For $L_1=L_2$ one gets
\begin{equation}
R_Z^*  = 0.372884880824589... \; .
\end{equation}
The estimate of the inverse critical temperature converges as 
\begin{equation}
 \bar{\beta}_{c,2D}(L_0,L)-\beta_{c,2D}(L_0)  = c(L_0) L^{-1/\nu_{2D}-\omega} + ... \;\;,
\end{equation}
where $\nu_{2D}=1$ is the critical exponent of the correlation length of the two-dimensional
Ising universality class.
In the case of $\xi_{2nd}/L$ and $U_4$ we have effectively $\omega=1.75$ 
due to the analytic background of the magnetic susceptibility. For 
$R_Z$ the leading correction is caused by the breaking of the rotational
symmetry by the lattice, resulting in $\omega=2$.  For a detailed discussion 
of corrections to scaling in two-dimensional Ising models see e.g. 
ref. \cite{our2DIsing}. Therefore, following ref. \cite{CaHa96}, we 
determine $\bar{\beta}_{c,2D}(L_0,L)$ by using the ratio $R_Z$ 
of partition functions. 

We determined the coefficients of the Taylor-expansion 
of the quantities we were interested in up to the third order around 
the inverse temperature $\beta_s$, where we simulated at. We have
chosen $\beta_s$ as good approximation of $\bar{\beta}_{c,2D}(L_0,L)$. 
This estimate is obtained by preliminary simulations, or from 
results for smaller lattice sizes that we had simulated already.
We solved eq.~(\ref{determinebc})  by replacing $R(\beta,L_0,L)$ 
on the left side of the equation by its third order Taylor-expansion 
around $\beta_s$.  

We simulated films of a thickness up to $L_0=64$ and $L=1024$.  In most cases
we performed $10^6$ update cycles. One cycle consists of one heat-bath sweep, 
one Todo-Suwa \cite{ToSu13} sweep,
a Swendsen-Wang \cite{SwWa87} cluster update and a wall-cluster \cite{wall} update plus a 
measurement of $Z_a/Z_p$ for each of the two directions. In total,
these simulations took about $2$ years of CPU time on a single core of a 
Quad-Core AMD Opteron(tm) 2378 CPU.

In table \ref{betac8} we give the results obtained for $L_0=4$  and $8$ for 
a large range of $L$.  Here we performed $10^8$ update cycles, except for $L_0=4$,
$L=256$ were we performed $3.3 \times 10^7$ update cycles, and $L_0=8$, $L=128$ and 
$L=512$ were we performed $5.5 \times 10^7$ and $9.6 \times 10^6$ update cycles, 
respectively. Fitting the data with the Ansatz
\begin{equation}
 \bar{\beta}_{c,2D}(L_0,L) = \beta_{c,2D}(L_0) + c L^{-3} 
\end{equation}
we get, taking all data into account, $\beta_{c,2D}(4) = 0.43968710(12)$, $c=-0.080(1)$ and
$\chi^2/$d.o.f.$=1.16$, and 
$\beta_{c,2D}(8) = 0.40724561(9)$, $c=-0.181(4)$ and $\chi^2/$d.o.f.$=1.39$ for $L_0=4$ and $8$, 
respectively.  Note that for $L_0=4$ and $8$ for $L \ge 16 L_0$ the estimate of $\bar{\beta}_{c,2D}(L_0,L)$ 
is consistent with $\beta_{c,2D}(L_0)$ within the statistical error. Therefore in the following, for other 
thicknesses $L_0$ we took $\bar{\beta}_{c,2D}(L_0,L)$ with $L \gtrapprox 16 L_0$ as our final 
estimate of $\beta_{c,2D}(L_0)$.

In order to match the reduced temperature of the two-dimensional Ising model and the
reduced temperature of the film, the derivative of $R_Z$ with respect to the reduced temperature $t$ 
at $R_Z^*$ is a useful quantity. Taking $\partial R_Z/\partial t = - \partial R_Z/\partial \beta$ at $R_Z^*$
means that the derivative is taken at $\bar{\beta}$, which is the solution of 
eq.~(\ref{determinebc}). It behaves as
\begin{equation}
\label{slope}
\bar{S}:= - \left . \frac{\partial R_Z}{\partial \beta} \right|_{R_Z=R_Z^*} = a L^{1/\nu_{2D}} \;\; (1 + c L^{-\omega} + ...) \;.
\end{equation}
In the fourth column of table \ref{betac8} we give 
$\bar{S}/L$ for $L_0=4$ and $8$ for all $L$ we have simulated. We fitted these data with the 
Ansatz
\begin{equation}
 \bar{S}/L = a + b L^{-2}  \;\;.
\end{equation}
Taking all data for $L_0=4$ into account we get $a=2.52502(18)$, $b=4.546(22)$, and $\chi^2/$d.o.f.$=1.17$,
while fitting all data for $L_0=8$ we get $a=3.8708(4)$, $b=18.91(19)$, and  $\chi^2/$d.o.f.$=0.68$.
In the case of $L_0=8$ we find that $\bar{S}/L$ for $L=128$ and $512$ is consistent with the asymptotic 
result obtained from the fit. For $L_0=4$ this is the case only for $L=128$ and $256$. For $L=64$ we see
a deviation of  about two standard deviations.
In table \ref{betac2Dall} we give our final estimates of $\beta_{c,2D}$ and the slope $\bar{S}/L$ 
for all thicknesses $L_0$ that we have simulated. We took results obtained for 
$L \gtrapprox 16 L_0$ as our final estimate. Note that for other values
of $L_0$ the statistics is considerably smaller and therefore the statistical errors larger
than for $L_0=4$ and $8$.

\begin{table}
\caption{\sl \label{betac8}
Numerical results for $\bar{\beta}_{c,2D}(L_0,L)$, eq.~(\ref{determinebc}), 
and the slope over the linear lattice size $\bar{S}/L$, eq.~(\ref{slope}), for the thicknesses
$L_0=4$ and $8$ for a large range of transversal lattice sizes $L$.
}
\begin{center}
\begin{tabular}{crll}
\hline
   \mc{1}{c}{$L_0$} & \mc{1}{c}{$L$}   &  \mc{1}{c}{$\bar{\beta}_{c,2D}$} &
  \mc{1}{c}{$- \frac{1}{L} \left . \frac{\partial R_Z}{\partial \beta} \right|_{R_Z=R_Z^*}$} \\
\hline
    4 &   8 & 0.4395281(25)  & 2.45398(26) \\  
    4 &  12 & 0.4396433(18)  & 2.49361(30) \\ 
    4 &  16 & 0.4396701(14)  & 2.50712(33) \\
    4 &  24 & 0.4396820(10)  & 2.51720(38) \\ 
    4 &  32 & 0.43968400(72) & 2.51981(41) \\ 
    4 &  48 & 0.43968644(50) & 2.52385(45) \\  
    4 &  64 & 0.43968672(40) & 2.52391(48) \\
    4 & 128 & 0.43968708(20) & 2.52453(54) \\ 
    4 & 256 & 0.43968704(18) & 2.5259(11) \\
\hline
    8 &  16 & 0.4072021(10)  & 3.7970(5) \\
    8 &  24 & 0.40723203(69) & 3.8380(6) \\
    8 &  32 & 0.40723991(53) & 3.8520(6) \\
    8 &  48 & 0.40724338(37) & 3.8621(7) \\ 
    8 &  64 & 0.40724454(27) & 3.8664(8) \\ 
    8 & 128 & 0.40724568(20) & 3.8710(12)\\ 
    8 & 512 & 0.40724571(12) & 3.8750(38)\\  
\hline
\end{tabular}
\end{center}
\end{table}

\begin{table}
\caption{\sl \label{betac2Dall}
Numerical results for the phase transition of films with $(O,O)$ boundary conditions. 
The thickness of the film is given by $L_0$ and $L$ is the linear extension in the 
two transversal directions. In the third column we give our estimate of the inverse
of the transition temperature $\beta_{c,2D}(L_0)$ as defined by eq.~(\ref{determinebc}).
In the fourth column we give $\bar{S}/L$ as defined by eq.~(\ref{slope}).
}
\begin{center}
\begin{tabular}{rrlr}
\hline
\mc{1}{c}{$L_0$} & \mc{1}{c}{$L$} &   \mc{1}{c}{$\beta_{c,2D}$}  & 
\mc{1}{c}{$ - \frac{1}{L} \left . \frac{\partial R_Z}{\partial \beta} \right|_{R_Z=R_Z^*}$} \\
\hline
    4 & 256 &  0.43968704(18) &2.5259(11) \\  
    5 & 160 &  0.4258884(15)  &   2.903(7)\phantom{00} \\  
    6 & 384 &  0.41724094(59) &   3.256(9)\phantom{00} \\
    7 & 112 &  0.4114039(17)  &   3.579(8)\phantom{00} \\
    8 & 512 &  0.40724571(12) &   3.875(4)\phantom{00} \\
    9 & 300 &  0.40416349(61) &   4.157(11)\phantom{0}  \\  
   10 & 256 &  0.40180434(69) &   4.430(12)\phantom{0}  \\ 
   11 & 256 &  0.39995347(66) &   4.669(13)\phantom{0}  \\
   12 & 192 &  0.39846789(82) &   4.918(13)\phantom{0}  \\
   13 & 192 &  0.39725856(81) &   5.147(14)\phantom{0}  \\
   14 & 256 &  0.39625624(59) &   5.391(15)\phantom{0}  \\ 
   15 & 256 &  0.39541461(57) &   5.568(16)\phantom{0}  \\
   16 & 256 &  0.39470035(55) &   5.789(16)\phantom{0}  \\ 
   17 & 256 &  0.39408852(54) &   6.048(17)\phantom{0}  \\
   24 & 384 &  0.39148514(31) &   7.350(23)\phantom{0}  \\
   25 & 384 &  0.39125639(31) &   7.524(24)\phantom{0}  \\ 
   32 & 512 &  0.39013763(21) &   8.661(29)\phantom{0}  \\
   48 & 768 &  0.38900912(12) &  10.988(46)\phantom{0}  \\ 
   64 &1024 &  0.38854284(8)  &  12.973(52)\phantom{0}  \\
\hline
\end{tabular}
\end{center}
\end{table}

The transition temperature of the film approaches the transition temperature
of the three-dimensional bulk system as the thickness $L_0$ of the film 
increases. Based on standard RG-arguments one expects \cite{Fi71,CaFi76}
\begin{equation}
\label{Fisherbc}
 \beta_{2D,c}(L_0) - \beta_{3D,c} \simeq a L_0^{-1/\nu}  \;\;.
\end{equation}
It turns out that corrections to scaling have to be included to 
fit our data. First we allowed for an effective thickness of the 
film
\begin{equation}
\label{Fisherfit1}
\beta_{c,2D}(L_0) - \beta_{c,3D} = a [L_0+L_s]^{-1/\nu} \;\;
\end{equation}
where we fixed $\beta_{c,3D}=0.387721735$ and $\nu=0.63002$.
The parameters of the fit are $a$ and $L_s$. Taking into account only 
thicknesses $L_{0} \ge 24$ we still get $\chi^2$/d.o.f. $=2.91$. 
Therefore we added a term that takes into account the leading 
analytic correction
\begin{equation}
\label{Fisherfit}
 \beta_{c,2D}(L_0) - \beta_{c,3D} = a [L_0+L_s]^{-1/\nu} + b [L_0+L_s]^{-2/\nu}
\end{equation}
where now $b$ is an additional parameter of the fit.
We find that already for $L_{0,min}=8$, where all data for $L_0 \ge L_{0,min}$
are taken into account, $\chi^2$/d.o.f. $\approx 1$. Hence the 
Ansatz~(\ref{Fisherfit}) along with the numerical values of the parameters
given in table \ref{betac2Dfit} can be used to obtain estimates of 
$\beta_{c,2D}(L_0)$ for thicknesses $8 \le L_0 \le 64$, where we have not 
simulated at. One should note that the parameters have a clear dependence 
on the value of $\nu$ that is used. For example fixing
$\nu=0.62992$ we get for $L_{0,min}=10$ the results $a=0.61875(8)$, $L_s=0.9569(48)$,
$0.4931(17)$ and $\chi^2$/d.o.f. $=0.83$.  An important observation is 
that the results obtained for $L_s$ are fully consistent with $L_s=0.96(2)$ 
obtained in ref. \cite{mycrossing} by studying the magnetization 
profile of films with $(O,+)$ boundary conditions at the critical point.

In terms of the scaling variable we get 
\begin{equation}
x_c= - a \xi_0^{-1/\nu} = - 6.444(10)
\end{equation}
where we have taken into account the uncertainties of $\nu$ and $\beta_c$.

In ref. \cite{KiOhIt96}  the authors computed $\beta_{c,2D}$ for the Ising model on 
the simple cubic lattice, using the crossing of the Binder cumulant. They obtain
$\beta_{c,2D} = 0.25844(4)$, $0.24289(3)$, $0.23587(2)$, $0.23209(3)$, $0.22965(3)$, and $0.22804(3)$
for the thicknesses $L_0 =4$, $6$, $8$, $10$, $12$ and $14$, respectively. 
In the case of the Ising model, we expect that corrections proportional to $L_0^{-\omega}$ with 
$\omega=0.832(6)$ contribute significantly, making the extrapolation to $L_0 \rightarrow \infty$ more
difficult than in the case of the improved Blume-Capel model. Despite this fact, to get at least a rough 
answer, we fitted the Ising data with the Ansatz~(\ref{Fisherfit1}), using 
$\beta_{c,3D}=0.22165462(2)$, see eq.~(A2) of \cite{MHcorrections}. We find $a=0.480(4)$, $L_s=1.18(5)$
and $\chi^2/$d.o.f. $=0.95$ taking into account all data for $L_0 \ge 8$.  Using the estimate of 
$\xi_0$ given in eq.~(A10) of \cite{MHcorrections} we get $x_c = - 6.37(5)$, which is close
with our estimate obtained for the improved Blume-Capel model. Eq.~(12) of
ref. \cite{Dohm14} gives $x_c \approx -6.5$ for the Ising universality class, which is in excellent 
agreement with our result.

\begin{table}
\caption{\sl \label{betac2Dfit}
Fitting the data of table \ref{betac2Dall} with the Ansatz~(\ref{Fisherfit}),
where $\beta_{c,bulk}=0.387721735$ and $\nu=0.63002$ are fixed, while 
$a$, $b$ and $L_s$ are the parameters of the fit. 
Data for thicknesses $L_0 \ge L_{0,min}$ are taken into account.
}
\begin{center}
\begin{tabular}{rllll}
\hline
\mc{1}{c}{$L_{0,min}$}& \mc{1}{c}{$a$} & \mc{1}{c}{$b$} & \mc{1}{c}{$L_s$} & \mc{1}{c}{$\chi^2$/d.o.f.} \\
\hline   
      6        &   0.61841(5) &   0.546(5) & 0.9665(18) & 3.41 \\
      7        &   0.61815(7) &   0.505(9) & 0.9570(29) & 1.23 \\
      8        &   0.61813(7) &   0.502(9)& 0.9560(29) & 1.12 \\
      9        &   0.61806(7) &   0.485(12)& 0.9513(37) & 0.97 \\
     10        &   0.61799(8) &   0.467(17)& 0.9467(46) & 0.85 \\
     11        &   0.61797(10)&   0.462(22)& 0.9453(61) & 0.93 \\
     12        &   0.61791(11)&   0.440(30)& 0.9401(77) & 0.91 \\
\hline
\end{tabular}
\end{center}
\end{table}

Finally we studied the behavior of $\bar{S}/L$ at 
the critical point as a function of the thickness $L_0$ of the film. It behaves as
\begin{equation}
 \bar{S}/L \simeq a \; [L_0+L_s]^{1/\nu-1 } \;\;.
\end{equation}
Performing various fits, using $L_s=0.96(2)$ and $\nu=0.63002(10)$ as input, 
we arrive at $a=1.12(1)$. In terms of the scaling variable $x=t [(L_0+L_s)/\xi_0]^{1/\nu}$
this means
\begin{equation}
 \bar{S}_x :=\frac{\partial R_Z}{\partial x} =  \bar{S} [(L_0+L_s)/\xi_0]^{-1/\nu} L
             \simeq a \xi_0^{1/\nu} \frac{L}{L_0+L_s}  = 0.1074(10) \frac{L}{L_0+L_s} \;.
\end{equation}
For the transversal correlation length of the film in the high 
temperature phase, eq.~(\ref{xi2DIsing})  translates to
\begin{equation}
\label{Filmxi}
 \xi_{Film} \simeq 1.99(2) [L_0 + L_s]  (x-x_c)^{-1}  \;\;
\end{equation} 
using
\begin{equation}
\label{slopeIsing}
\lim_{L \rightarrow \infty} [\bar{S}/L]_{2D Ising} = \lim_{L \rightarrow \infty} \; \frac{1}{L} \left . \frac{\partial Z_a/Z_p}{\partial \tau} \right|_{\tau=0}   =  0.3021247100407...  \;\;
\end{equation}
for the two-dimensional Ising model.

\begin{figure}
\begin{center}
\includegraphics[width=14.5cm]{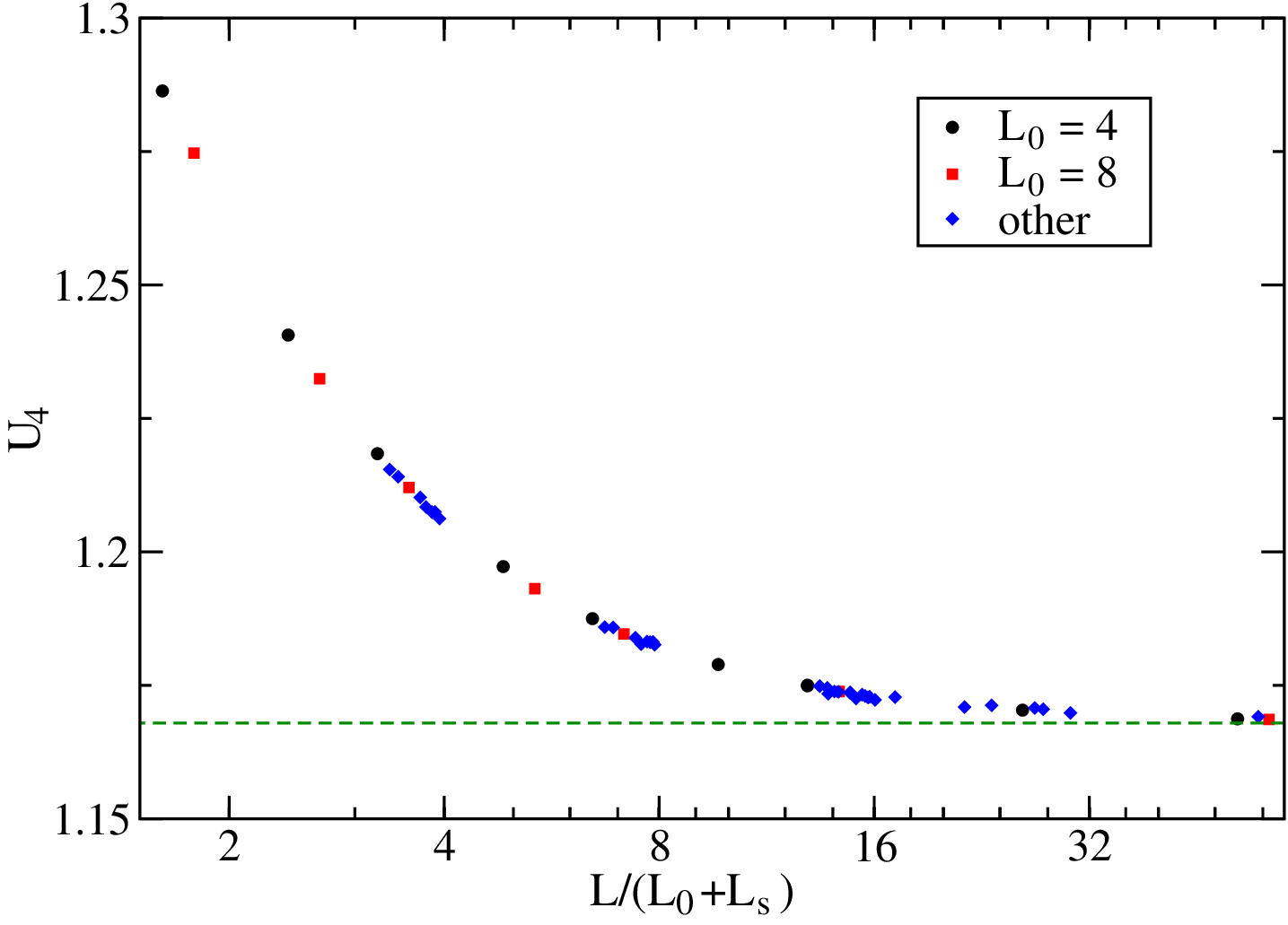}
\caption{\label{binderplot}
We plot $\bar{U}_4$ as a function of $L/(L_0+L_s)$ with $L_s=0.96$. For comparison we
give $U_4^*=1.1679229$ as green dashed line. The data points for $L_0=4$ and $8$ are given as black circles
and red squares, respectively.  For all other thicknesses, the data points are shown as blue diamonds.
The error bars are smaller than the size of the symbols.
For the definition of the quantities and a discussion see the text.
}
\end{center}
\end{figure}

In Fig. \ref{binderplot} we plot $\bar{U}_4$ as a function of $L/(L_0+L_s)$, 
where $\bar{U}_4$ is the Binder cumulant $U_4=\frac{\langle m^4 \rangle}{\langle m^2 \rangle^2}$
at $R_Z=R_Z^*$, where $m=\sum_x s_x$ is the magnetization. Following ref. \cite{SaSo00}
$U_4^*=1.1679229\pm 0.0000047$. 
Its interesting to see that already starting from $L_0=4$, finite $L$ effects nicely scale with 
the effective thickness $L_0+L_s$. We have checked that the decay of corrections with increasing
$L$ is consistent with  $\bar{U}_4-U_4^* \propto L^{-7/4}$, as theoretically expected.
Finally we convinced ourself that $\bar{\xi}_{2nd}/L$ converges to 
$(\xi_{2nd}/L)=0.9050488292 \pm 0.0000000004$ \cite{SaSo00}  as $L/(L_0+L_s) \rightarrow \infty$.
These observations strongly support the hypothesis that the transition of the film, for any thickness
$L_0$, belongs  to the two-dimensional Ising universality class.

\subsection{Thermodynamic Casimir force for $(O,O)$ boundary conditions}
\label{casimiroo}
The thermodynamic Casimir force for  $(O,O)$ boundary conditions has been 
studied for the Ising model \cite{VaGaMaDi08} and the improved 
Blume-Capel model \cite{PaTrDi13}.

We have simulated  films of the thicknesses $L_0=8.5$, $12.5$, $16.5$ and $24.5$.
For the parameters of the update we took $n_{exc}=20$ throughout and $i_r=1$, $2$, $2$, and $3$ 
for $L_0=8.5$, $12.5$, $16.5$, and $24.5$, respectively.
We simulated the transversal lattices sizes $L=32$, $64$, $128$ and 
$256$ for $L_0=8.5$, $L=48$, $96$ and $192$ for $L_0=12.5$, $L=64$ for
$L_0=16.5$, and $L=96$ and $192$ for $L_0=24.5$. 
For all thicknesses  we simulated at slightly more than hundred values of $\beta$
in the neighborhood of the bulk critical point. The larger transversal 
lattices sizes were simulated at less values of $\beta$ than the 
smaller ones, focussing at the neighborhood of the transition of the 
film.
We performed $10^6$ update cycles for each value of $\beta$ and 
most lattice sizes. Exceptions are $(L_0,L)=(8.5,256)$ and $(12.5,192)$ 
were we performed only  $2 \times 10^5$ update cycles. In total we used
about 5 years of CPU time on a single core of an AMD Opteron 2378 
running at 2.4GHz. 

\begin{figure}
\begin{center}
\includegraphics[width=15cm]{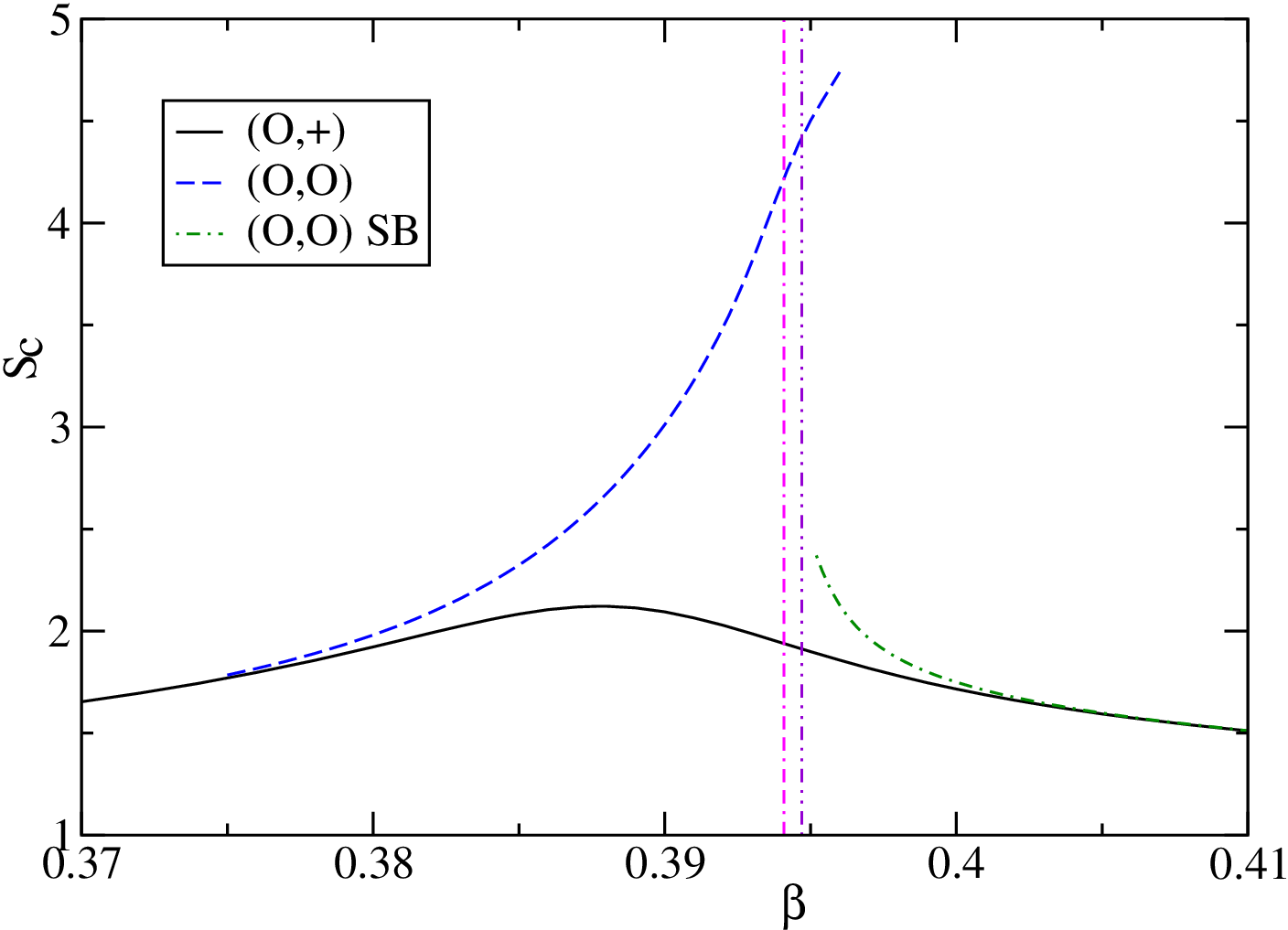}
\caption{\label{ballOO16}
We plot the size of the frozen exchange clusters $S_c$ for the thickness $L_0=16.5$.  We compare
$(O,+)$ and $(O,O)$ boundary conditions. In case of $(O,O)$ we give results for the simulation with
and without breaking of the $\mathbb{Z}_2$ symmetry. The vertical lines give the inverse transition temperature
of films of the thickness $L_0=16$ and $17$.
}
\end{center}
\end{figure}

\begin{figure}
\begin{center}
\includegraphics[width=15cm]{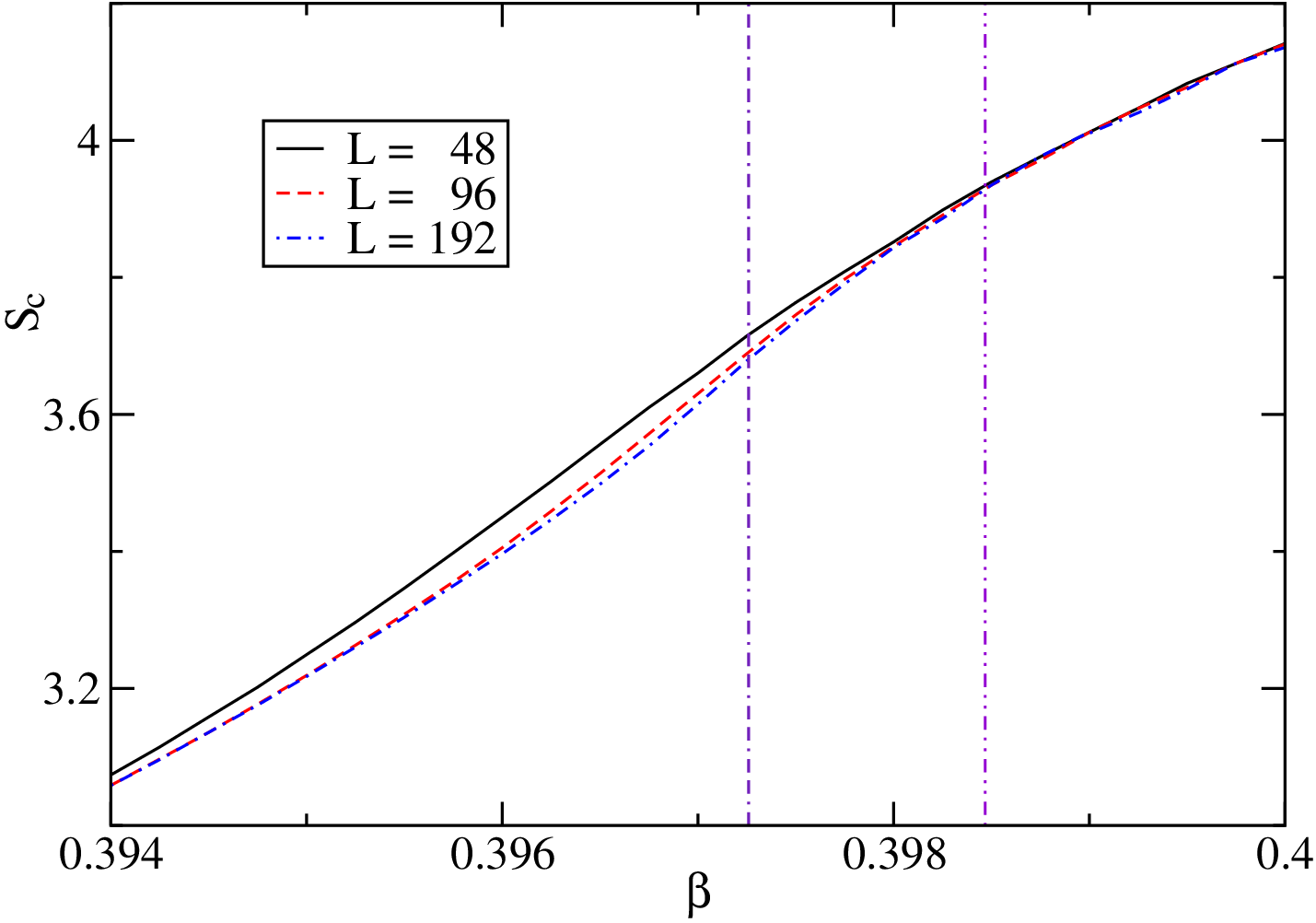}
\caption{\label{ball12fss}
We plot the size of the frozen exchange clusters $S_c$ for the thickness $L_0=12.5$ for the 
three transversal extensions $L=48$, $96$ and $192$. 
}
\end{center}
\end{figure}

Let us first discuss the performance of the exchange cluster algorithm.
In Fig.~\ref{ballOO16} we plot the average size per area of the frozen exchange clusters 
$S_c$ for $L_0=16.5$ and $L=64$.
For comparison we give our result for $(O,+)$ boundary conditions, 
where the exchange cluster update  is
performed at the $O$ boundary. For small $\beta$ the curves for $(O,+)$ and $(O,O)$ boundary 
conditions fall on top of each other. While for $(O,+)$ boundary conditions a maximum is reached 
at $\beta \approx  \beta_c$,
for  $(O,O)$ ones we find that $S_c$ is growing monotonically with increasing $\beta$. At the inverse
transition temperatures of the two films, $S_c$ is already a significant fraction of the thickness $L_0$ of the 
film.  We find $S_c \approx 3.25$, $3.82$, $4.32$, and $5.15$ 
at $\beta = (\beta_{c,2D}(L_0+1/2) + \beta_{c,2D}(L_0-1/2))/2$, for $L_0=8.5$, $12.5$, $16.5$, and $24.5$, 
respectively. For those thicknesses, where we have simulated more than one value of $L$, we find at 
$\beta_{c,2D}$ and in a certain neighborhood below a small dependence of $S_c$ on $L$. 
In Fig.~\ref{ball12fss}, we plot as an example $S_c$ for $L_0=12.5$ and 
$L=48$, $96$ and $192$. 

Looking at the simulation in the low temperature phase in detail we find that the large frozen exchange 
clusters grow, when the magnetization of the two systems have different sign. Physically one
could force the two systems to have the same sign by applying a bulk field $h$, such that 
$h L_0 L^2 m \gg 1$, where $m$ is the magnetization of the film. The larger $L$, the smaller the 
amplitude of the bulk field $h$
could be chosen. At the end one would extrapolate the results obtained to $h=0$.  Here instead, we 
break the symmetry by hand. After the sweeps with the heat-bath and the Todo-Suwa algorithm and the 
Swendsen-Wang cluster update of the two systems, before starting the $n_{exc}$ exchange cluster updates, 
we forced the two systems to positive or zero magnetization. To this end, we multiplied all spins of a system
with $-1$, if its magnetization is negative. This is certainly an update of the configuration that does not
fulfil balance and hence we introduce a systematic error. However, we expect that this error vanishes 
in the limit $L \rightarrow \infty$ and also decreases as we go deeper into the symmetry broken phase.
In Fig.~\ref{ballOO16} we also give $S_c$ for simulations with this explicit symmetry breaking (SB). 
We find that indeed $S_c$ is much smaller than for the simulation without SB. Also in the low
temperature phase of the films, $S_c$ is now decreasing with increasing $\beta$. For large $\beta$, the 
curve is falling on top of that for $(O,+)$ boundary conditions.

Let us briefly discuss the gain~(\ref{gaindef}) that we do not plot here.
Without SB, for all $L_0$ that we studied, it is almost linearly decreasing 
with increasing $\beta$,
until $\beta_{c,2D}(L_0+1/2)$ is reached. Starting from this point it stays roughly
constant with a value that is approximately equal to $1.4$. For 
$\beta \approx 0.389$ gain takes about the same value $4$ for all thicknesses
that we study. Using SB, starting from $\beta$ above $\beta_{c,2D}(L_0-1/2)$,
the gain rapidly increases with increasing $\beta$. For example the gain 
reaches the value $5$ at $\beta \approx 0.421$, $0.403$, $0.3973$ and $0.3925$ for 
$L_0=8.5$, $12.5$, $16.5$ and $24.5$, respectively.

For $\beta$ somewhat larger than $\beta_{c,2D}$ we simulated with SB and without.
For example for $L_0=8.5$ we find that the results for $\Delta E$ are consistent
at the level of our statistical accuracy starting from $\beta=0.409$, $0.408$, 
$0.4075$, and $0.407$ for $L=32$, $64$, $128$ and $256$, respectively. In our
analysis of the thermodynamic Casimir force below,
we have used the results obtained with SB starting from slightly larger
values of $\beta$, to have a safety margin.

\begin{figure}
\begin{center}
\includegraphics[width=15cm]{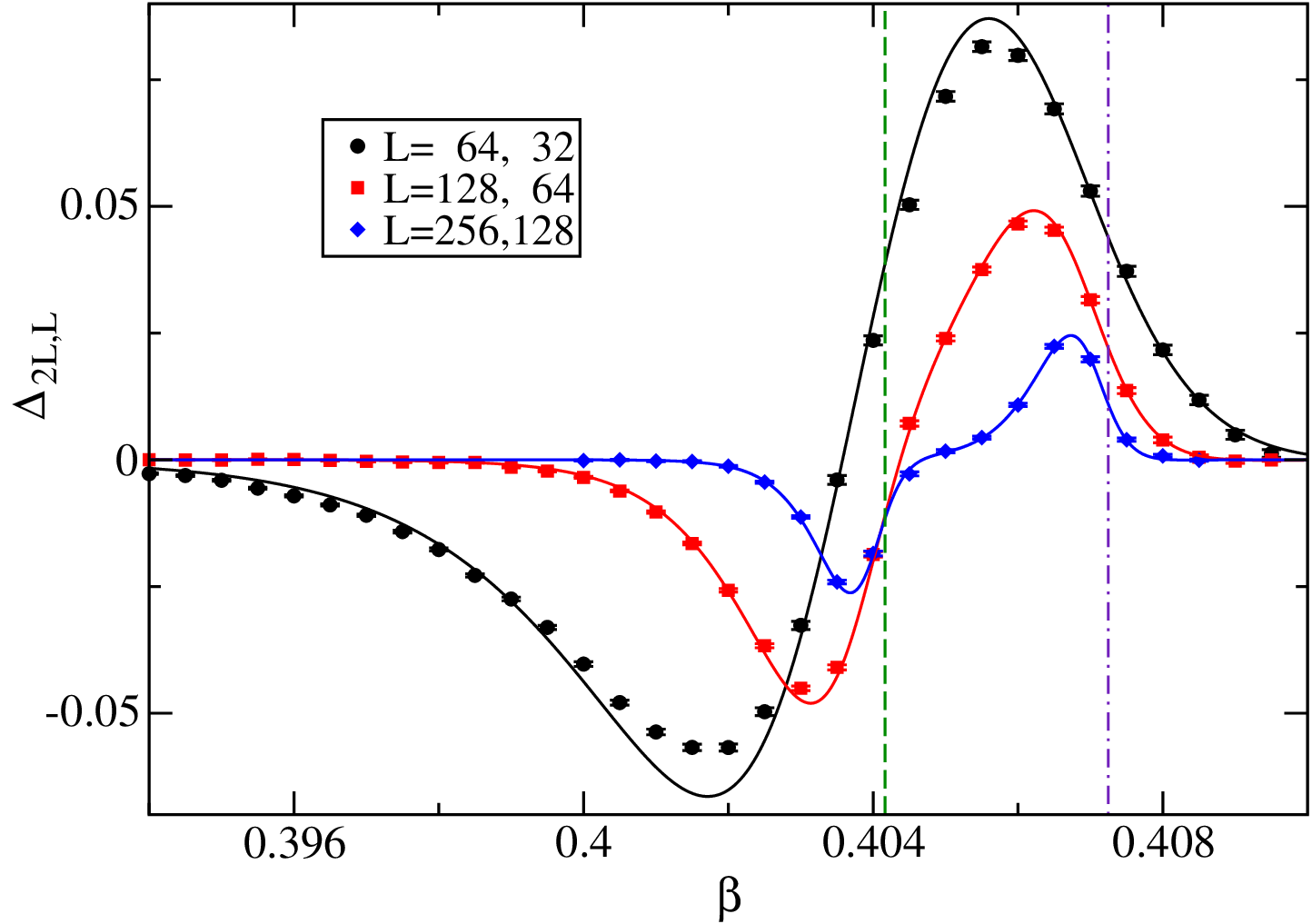}
\caption{\label{diffs8}
We plot $\Delta_{2L,L} = \Delta E(L_0,2 L) - \Delta E(L_0,L)$ for $L_0=8.5$.
Our numerical data are given by black circles, red squares and blue diamonds for and $L=32$, $64$ and 
$128$, respectively. The solid lines give the theoretical prediction, obtained from the universal 
finite size scaling function of the free energy density of the 2D Ising  transition.
The vertical dashed green line indicates the phase transition for $L_0=9$ and the vertical dashed-dotted
violet line the phase transition for $L_0=8$.
}
\end{center}
\end{figure}

In a first step of the analysis we check whether finite $L$ effects in $\Delta E_{ex}$ are well
described by the universal finite size scaling function $g_{n}$, eq.~(\ref{defgn}).
In Fig.~\ref{diffs8} we plot $\Delta_{2L,L} = \Delta E(L_0,2 L) - \Delta E(L_0,L)$   for $L_0 = 8.5$ and 
$L=32$, $64$ and $128$. Note that $\Delta E(L_0,2 L) - \Delta E(L_0,L)=\Delta E_{ex}(L_0,2 L) - \Delta E_{ex}(L_0,L)$,
since the bulk energy density cancels.
Our numerical results are compared with the prediction obtained from the universal finite
size scaling function $g_2$. As input we use the inverse transition temperature $\beta_{c,2D}$ and 
the slope of $R_Z$ at $R_Z^*$ given in table \ref{betac2Dall}, and eq.~(\ref{slopeIsing}):
\begin{equation}
\label{finiteE2} 
[E(L_0,2 L) - E(L_0,L)]_{predict}=  
 - \frac{\mbox{d}}{\mbox{d} \beta}   g_2(c \; [\beta_{c,2D}(L_0)-\beta] \; L) \; L^{-2} 
\end{equation}
where 
\begin{equation}
c =\frac{[\bar{S}/L]_{Film,L_0}} {[\bar{S}/L]_{2D Ising}} \;\;.
\end{equation}

We find that for $L=32$ the data are quite close to the prediction obtained from the universal
finite size scaling function $g_2$. Note that for $(L_0,L)=(12.5,48)$ and $(24.5,96)$ 
similar observations can be made. Going to $L=64$ the matching 
between the data points and the predicted behavior becomes better. Only at the minimum and the 
maximum of the curve a small missmatch can be observed. For $L_0=12.5$ and $L=96$ a similar 
observation can be made. Finally, for $L=128$, at the
level of our statistical accuracy, the match between the data points and the predicted behavior is 
perfect.

\begin{figure}
\begin{center}
\includegraphics[width=15cm]{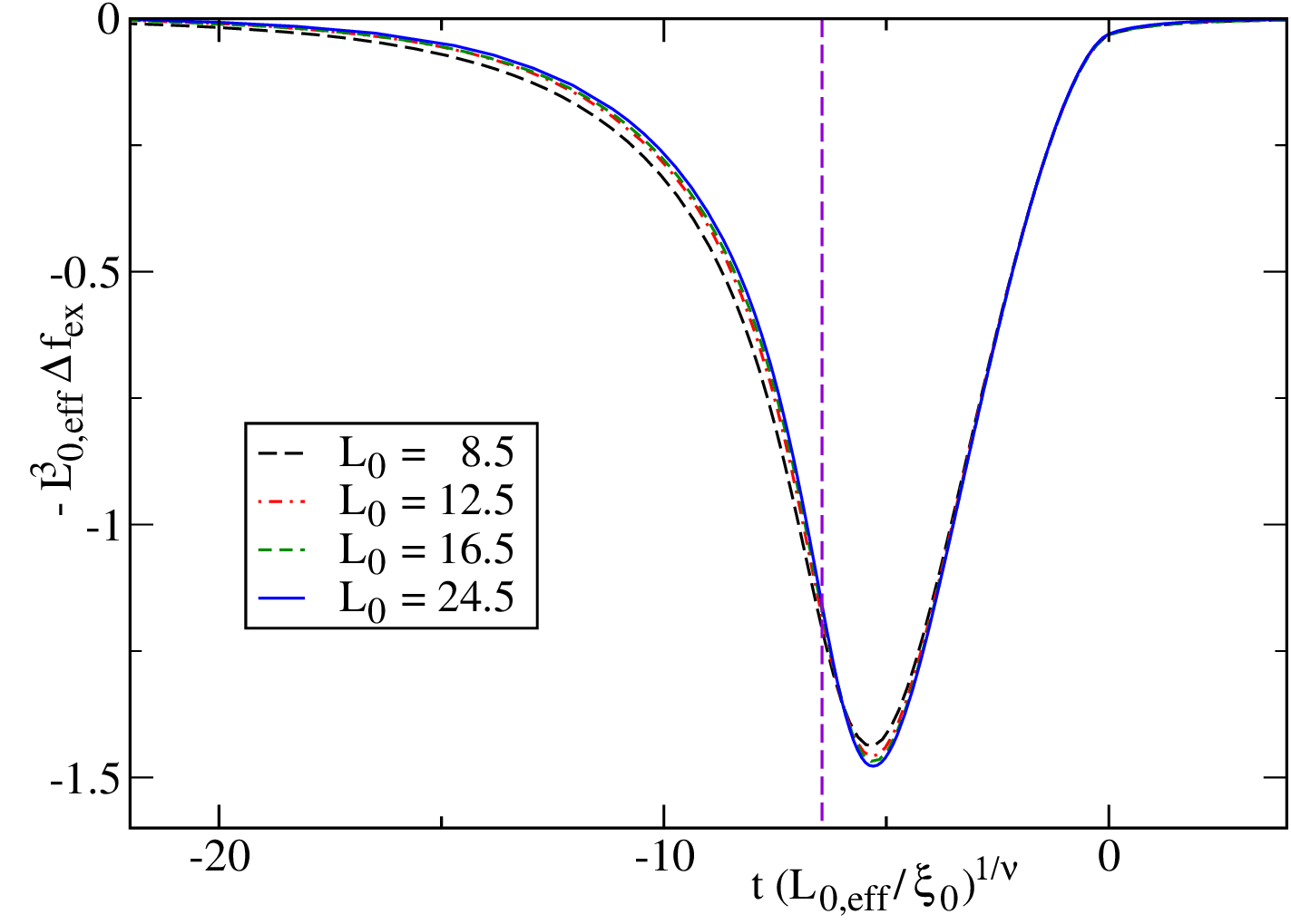}
\caption{\label{smallqplot}
We plot  $-L_{0,eff}^3 \Delta f_{ex}$ as function of $t (L_{0,eff}/\xi_0)^{1/\nu}$ for 
$(L_0,L)=(8.5,32)$, $(12.5,48)$, $(16.5,64)$, and $(24.5,96)$,  where we used $L_{0,eff}=L_0+L_s$ with
$L_s=0.96$, $\xi_0=0.2283$ and $\nu=0.63002$. 
The vertical dashed violet line gives the position of the phase transition of the film. 
}
\end{center}
\end{figure}

Next we checked how the results for the thermodynamic Casimir force are scaling with the 
thickness $L_0$ of the film. To this end we plot in Fig.~\ref{smallqplot} our numerical results for
$-(L_0+L_s)^3 \Delta f_{ex}$ as function of $t [(L_0+L_s)/\xi_0]^{1/\nu}$ for
$(L_0,L)=(8.5,32)$, $(12.5,48)$, $(16.5,64)$, and $(24.5,96)$.
Since $L/[L_0+L_s]$ is similar for these lattices, we expect that finite $L/[L_0+L_s]$ corrections
to scaling are similar.
For $x \gtrapprox -3$ the curves fall almost perfectly on top of each other. In contrast, for smaller
values of $x$ the different curves can be resolved at our level of numerical accuracy. In particular
the one for $L_0=8.5$ is clearly different from the others. Since the difference between the results
for $L_0=16.5$ and $24.5$ is rather minute, we expect that for $L_0=24.5$ deviations from the 
scaling limit are of a similar size as our statistical errors for  $L_0=24.5$.  A more quantitative
discussion of corrections will be given below, when we analyze the position of the minimum of the 
scaling function $\theta$.

\begin{figure}
\begin{center}
\includegraphics[width=15cm]{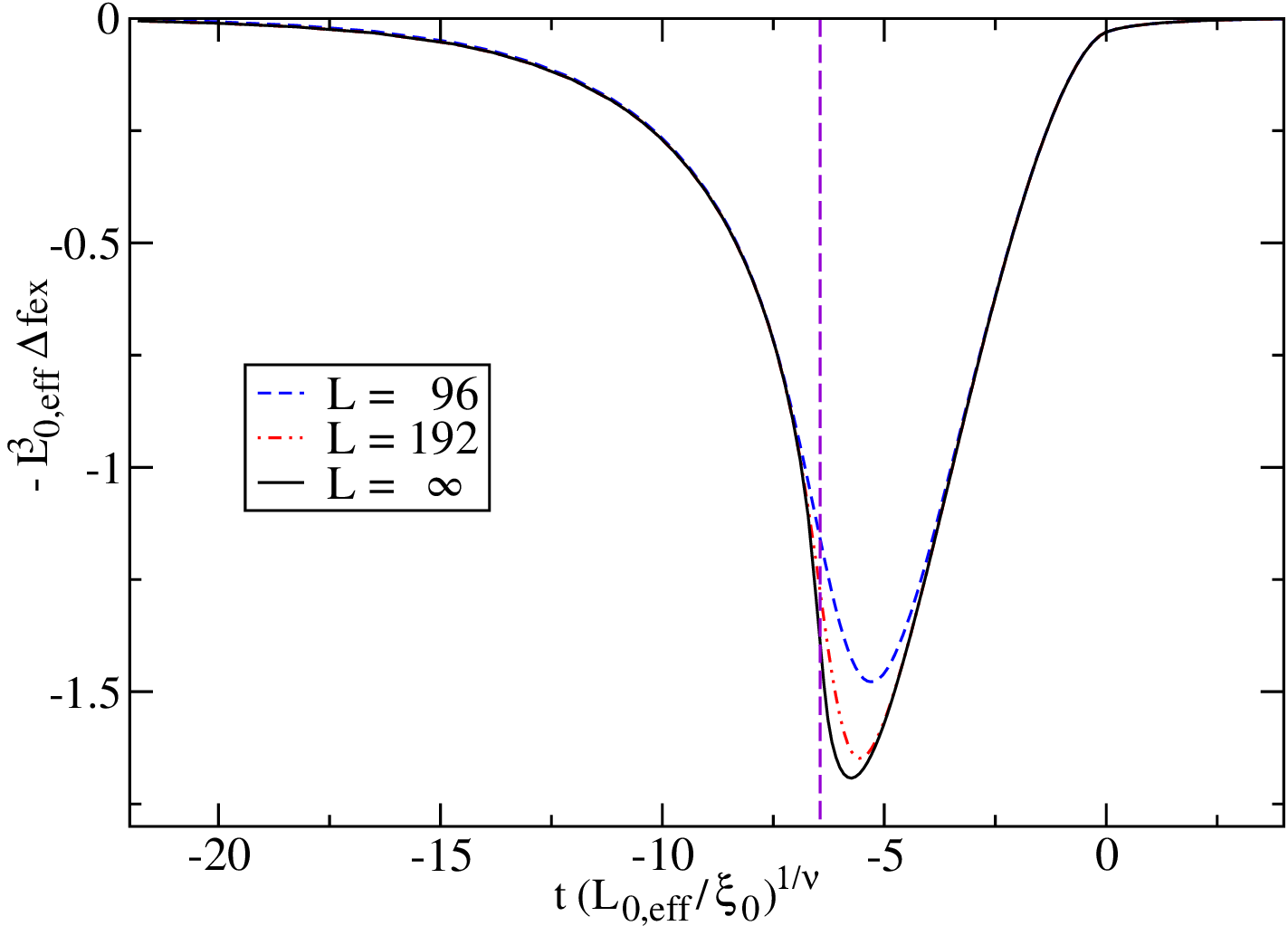}
\caption{\label{finalOO}
We plot  $-L_{0,eff}^3 \Delta f_{ex}$ as function of $t (L_{0,eff}/\xi_0)^{1/\nu}$ for
$L_0=24.5$ for $L=96$ and $192$ and our extrapolation to $L \rightarrow \infty$.
We used $L_{0,eff}=L_0+L_s$ with $L_s=0.96$, $\xi_0=0.2283$ and $\nu=0.63002$.
The vertical dashed violet line indicates $x_c$.
}
\end{center}
\end{figure}

Finally, in Fig.~\ref{finalOO} we plot 
$-(L_0+L_s)^3 \Delta f_{ex}$ as function of $t [(L_0+L_s)/\xi_0]^{1/\nu}$ for $L_0=24.5$ for 
$L=96$ and $192$ and our extrapolation of the $L=192$ result to $L \rightarrow \infty$ obtained by 
using the universal scaling 
function $g_{\infty}$. We see that the minimum of $\theta$ deepens as the lattice size increases and 
the position  of the minimum approaches $x_c$. The position of the minimum for $L \rightarrow \infty$
is close to $x_c$ but definitely different from it. We extrapolated our results obtained for $L_0=8.5$,
$L=256$ and $L_0=12.5$, $L=192$ to $L = \infty$. Note that for $L_0=16.5$ we have only data for $L=64$, and
therefore a reliable extrapolation is not possible.
Analyzing these data we find that $(x_{min}, \theta_{min})=$  $(-5.771(2)[19],-1.6922(4)[108])$, 
$(-5.757(5)[14],-1.6924(8)[76])$, and $(-5.746(7)[7],-1.6925(10)[40])$ for $L_0=8.5$, $12.5$ and $24.5$, 
respectively. Again the number in $[]$ gives the error due to the uncertainty of $L_s$.
As our final result for the limit $L_0 \rightarrow \infty$ we quote
\begin{equation}
 x_{min} = -5.75(2)  \;\;,\;\;\;   \theta_{(O,O)}(x_{min}) = -1.693(5)  \;\;
\end{equation}
which is consistent with the results obtained for the three different thicknesses.

Since $x_{min}$ is definitely larger than $x_{c}=-6.444(10)$, the correlation length of the film at $x_{min}$ is 
finite. Following eq.~(\ref{Filmxi}), $\xi_{Film}(x_{min}) \approx  1.99 \times (-5.75+6.444) \; L_{0,eff}$
$\approx 1.4 L_{0,eff}$.  For $L \gtrapprox 10 \xi_{Film}$,
finite $L$ effects should be small. Hence for $L \gtrapprox 14 \; L_0$ the features of the minimum of 
$\theta$ should be essentially independent of $L$. This is consistent with the observations of ref.
\cite{PaTrDi13}; See in particular their Fig.~16.  
Obviously, in an experiment no periodic boundary conditions 
can be applied. Still $\xi_{Film}(x_{min})$ indicates how large the transversal linear size of the 
system should be to avoid finite size effects.

Our result can be compared with ref. \cite{VaGaMaDi08} who simulated the Ising model
on the simple cubic lattice and the thicknesses $L_0=7.5$, $11.5$, $15.5$, and $19.5$.   Throughout, 
they used $\rho=L_0/L=1/6$.  They arrive at $(x_{min}, \theta_{min}) = (-5.74(2), -1.629(3))$
and $(-5.73(4),-1.41(1))$, depending on whether they use their eqs.~(18,20) or eq.~(21) to extrapolate
to $L_0 \rightarrow \infty$.  Interpolating our data to  $\rho=1/6$  using the 
universal finite size scaling function of the free energy, we arrive at $x_{min} \approx - 5.46$ and 
$\theta_{min} \approx -1.61$. Hence the apparently good agreement of $x_{min}$ with our result seems
to be a coincidence.
The authors of ref. \cite{PaTrDi13} give no explicit result for $x_{min}$ and $\theta_{min}$ in the text. From 
the insert of their Fig.~16 we read off $x_{min} \approx -5.5(1)$ and $\theta_{min} \approx -1.66(5)$.
The main reason for the larger error bar of \cite{PaTrDi13} compared with us is that they use $L_s=0.8(2)$,
$c'$ in their notation, instead of our $L_s=0.96(2)$.
Using field theoretic methods the author of ref. \cite{Dohm14} arrives at $x_{min} \approx -5.53$
and $\theta_{min} \approx - 1.5$. 

Similar to eq.~(\ref{generalizedFSS}),
the thermodynamic Casimir force per area as a function of the inverse temperature
$\beta$ and the surface fields $h_1$ and $h_2$ follows the scaling law
\begin{equation}
\label{generalizedFSSOO}
F_{Casimir}(\beta,h_1,h_2) =  k_B T L_0^{-d} \Theta_{(O,O)}(x, x_{h_1},x_{h_2})
\end{equation}
where
\begin{equation}
\label{defxh2}
 x_{h_1} = h_1 [L_0/l_{ex,nor,0}]^{y_{h_1}} \;\;,\;x_{h_2} = h_2 [L_0/l_{ex,nor,0}]^{y_{h_1}}
\end{equation}
where for our model $l_{ex,nor,0}=0.213(3)$, eq. (73) of \cite{mycrossing}, and the
surface critical RG-exponent $y_{h_1} = 0.7249(6)$ , eq. (52) of \cite{mycrossing}.

\begin{figure}
\begin{center}
\includegraphics[width=15cm]{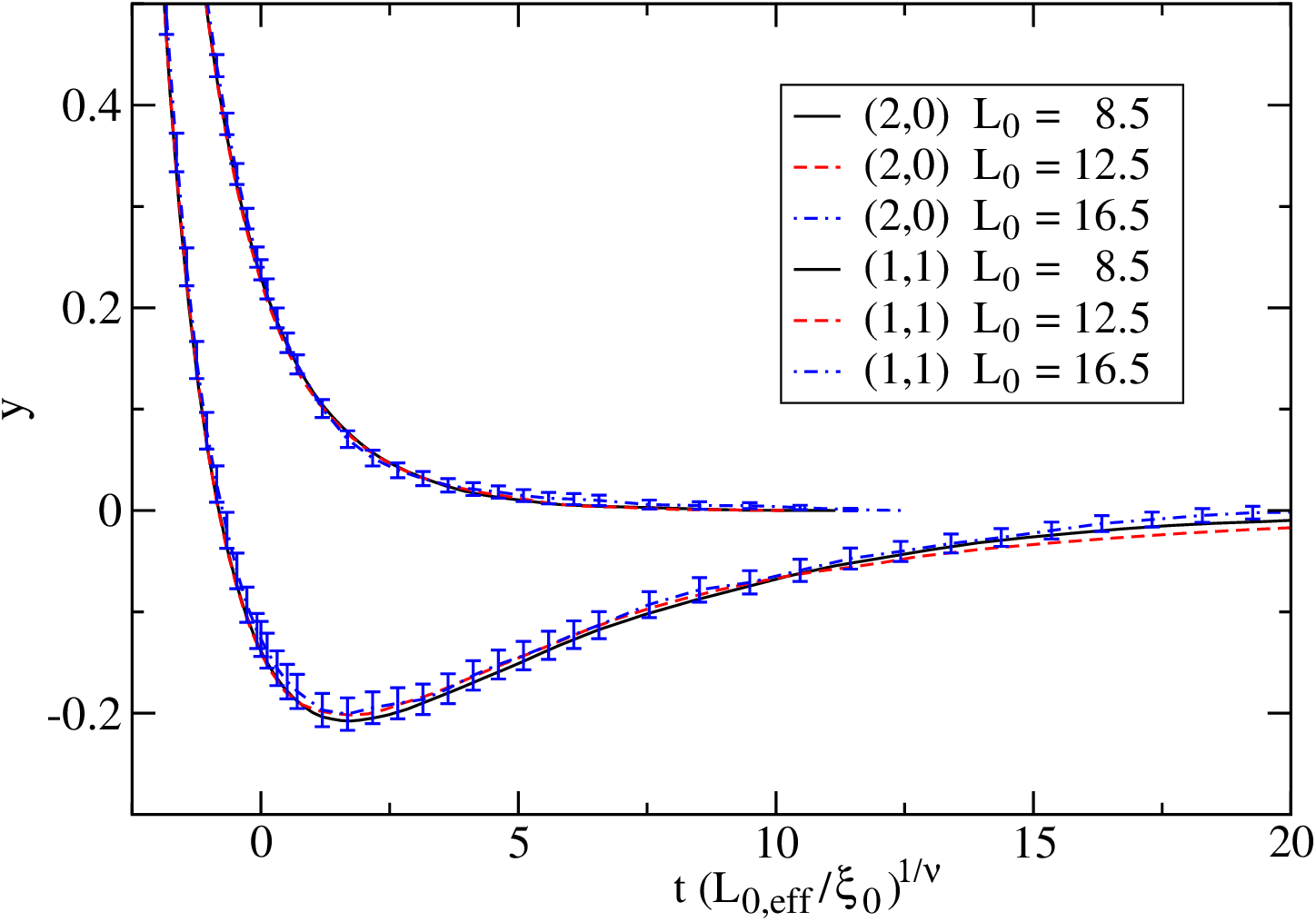}
\caption{\label{plotoohh}
We plot 
$y^{(2,0)}=  - L_{0,eff}^3 (L_{0,eff}/l_{ex,nor,0})^{-2 y_{h_1}}
 \frac{\partial^2 \Delta f_{ex}}{\partial h_1^2}$ 
and
$y^{(1,1)}=  - L_{0,eff}^3 (L_{0,eff}/l_{ex,nor,0})^{-2 y_{h_1}}
 \frac{\partial^2 \Delta f_{ex}}{\partial h_1 \partial h_2}$
at $h_1=h_2=0$
as
a function of $ t (L_{0,eff}/\xi_0)^{1/\nu}$ for $(O,O)$ boundary conditions
for the thicknesses $L_0=8.5$, $12.5$, and $16.5$.  To this end,
we have used $L_{0,eff}=L_0+L_s$ with $L_s=0.96$, $\xi_0=0.2283$,
$\nu=0.63002$, $l_{ex,nor,0}=0.213$, and $y_{h_1}=0.7249$. 
To keep the figure readable, error bars are only shown
for $L_0=16.5$, where they are the largest. We use the same types of lines 
for $y^{(2,0)}$ and $y^{(1,1)}$. Note that $y^{(1,1)} < y^{(2,0)}$ in the whole
range that is plotted.
}
\end{center}
\end{figure}

The partial derivatives of $\Delta f_{ex}$ with respect to $h_1$ and $h_2$
at $h_1=h_2=0$ are determined in a similar fashion as for $(O,+)$
boundary conditions. In the high temperature phase of the film,  
due to the $\mathbb{Z}_2$-symmetry of the problem, the first derivatives vanish.
In Fig.~\ref{plotoohh} we plot our results for
\begin{equation}
\theta^{(1,1)}(x) \equiv \left . \frac{\partial^2 \Theta(x,x_{h_1},x_{h_2})}{\partial x_{h_1} \partial x_{h_2}}
\right |_{x_{h_1}=x_{h_2}=0} \simeq - L_{0,eff}^3 (L_{0,eff}/l_{ex,nor,0})^{-2 y_{h_1}}
 \frac{\partial^2 \Delta f_{ex}}{\partial h_1 \partial h_2}
\end{equation}
and
\begin{equation}
\theta^{(2,0)}(x) \equiv \left . \frac{\partial^2 \Theta(x,x_{h_1},x_{h_2})}{\partial x_{h_1}^2 }
\right |_{x_{h_1}=x_{h_2}=0} \simeq - L_{0,eff}^3 (L_{0,eff}/l_{ex,nor,0})^{-2 y_{h_1}}
\frac{\partial^2 \Delta f_{ex}}{\partial h_1^2}  \;\;.
\end{equation}
Despite variance reduction, the statistical error increases rapidly with increasing thickness.
Our data for $L_0=24.5$ already have a quite large statistical error and we therefore did not
plot them in Fig.~\ref{plotoohh}. 
In the high temperature phase of the bulk system only $\theta^{(1,1)}$ has 
a significant amplitude and it is negative. Going to lower temperatures,
towards the transition temperature of the film, both $\theta^{(1,1)}$
and $\theta^{(2,0)}=\theta^{(0,2)}$ rapidly increase. Also $\theta^{(1,1)}$
and $\theta^{(2,0)}=\theta^{(0,2)}$ approach each other in this range.
As a result, in this range, the thermodynamic Casimir force varies much
less with $h_1$ for $h_1=-h_2$ than for example for $h_1=h_2$.

At the minimum of $\theta_{(O,O)}$ we have $\theta^{(2,0)} \approx \theta^{(1,1)}  \approx 500$. 
This means that for example for  $h_1 = h_2$, already for 
$x_{h_1} \gtrapprox 0.03$ the characteristics of the thermodynamic Casimir
force for $(O,O)$ boundary conditions are completely wiped out. 

\begin{figure}
\begin{center}
\includegraphics[width=15cm]{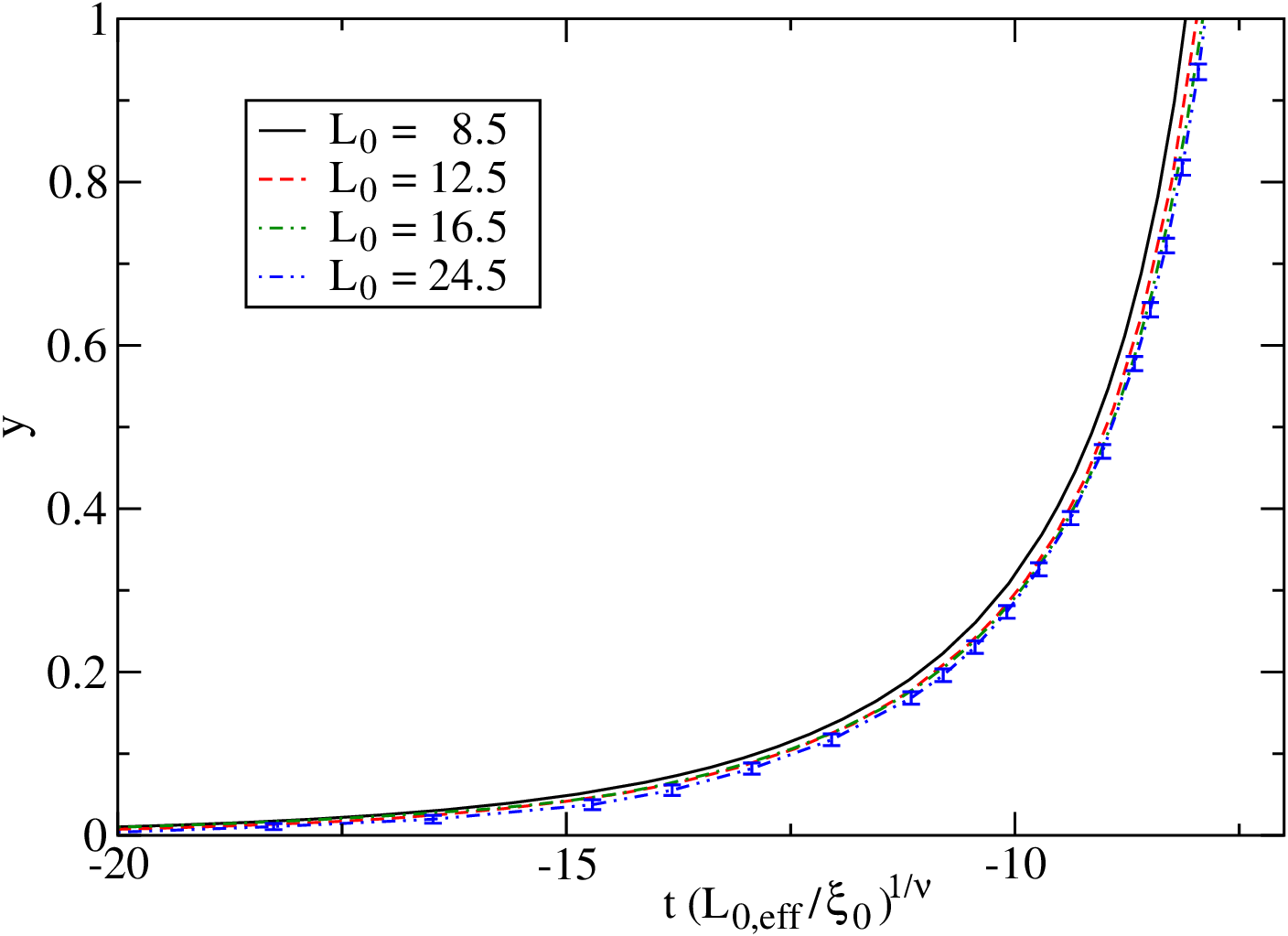}
\caption{\label{plotUX}
We plot
$y=  - L_{0,eff}^3 (L_{0,eff}/l_{ex,nor,0})^{-y_{h_1}}
 \frac{\partial \Delta f_{ex}}{\partial h_1}$ as
a function of $ t (L_{0,eff}/\xi_0)^{1/\nu}$ for $(0,O)$ boundary conditions
for the thicknesses $L_0=8.5$, $12.5$, $16.5$, and $24.5$ for the low
temperature phase of the film. To this end,
we have used $L_{0,eff}=L_0+L_s$ with $L_s=0.96$, $\xi_0=0.2283$,
$\nu = 0.63002$, $l_{ex,nor,0}=0.213$, and $y_{h_1}=0.7249$. 
}
\end{center}
\end{figure}

For completeness we also give our results for temperatures below the 
transition temperature of the film. Here we rely on our simulations with 
SB.  Since the $\mathbb{Z}_2$ symmetry is broken, the first derivative with 
respect to $x_{h_1}$ does not vanish. The numerical integration is started 
at large values of $\beta$. Our numerical data are plotted in Fig.~\ref{plotUX}.
For $L_0=12.5$, $16.5$ and $24.5$, we find a quite good collapse of the data 
on a single scaling curve. The function
$\theta'$ is positive in the whole range $x<x_c$. It rapidly increases as
$x_c$ is approached.

\begin{figure}
\begin{center}
\includegraphics[width=15cm]{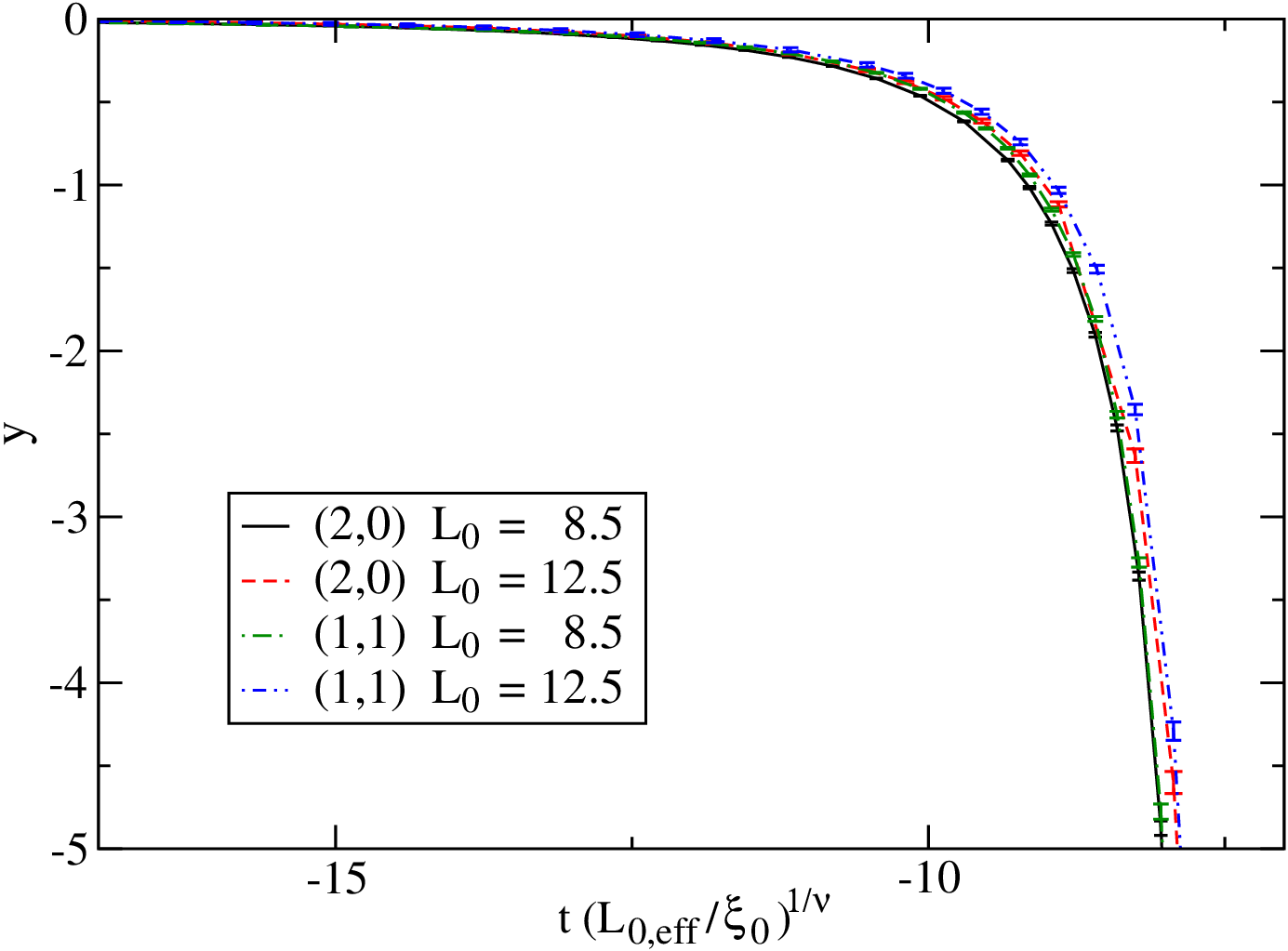}
\caption{\label{plotUXX}
We plot 
$y^{(2,0)}=  - L_{0,eff}^3 (L_{0,eff}/l_{ex,nor,0})^{-2 y_{h_1}}
 \frac{\partial^2 \Delta f_{ex}}{\partial h_1^2}$ 
and 
$y^{(1,1)}=  - L_{0,eff}^3 (L_{0,eff}/l_{ex,nor,0})^{-2 y_{h_1}}
 \frac{\partial^2 \Delta f_{ex}}{\partial h_1 \partial h_2}$ 
as
a function of $ t (L_{0,eff}/\xi_0)^{1/\nu}$ for $(O,O)$ boundary conditions
for the thicknesses $L_0=8.5$ and $12.5$ in the low temperature phase 
of the film.  To this end,
we have used $L_{0,eff}=L_0+L_s$ with $L_s=0.96$, $\xi_0=0.2283$,
$\nu=0.63002$, $l_{ex,nor,0}=0.213$, and $y_{h_1}=0.7249$. 
}
\end{center}
\end{figure}

Finally in Fig.~\ref{plotUXX} we plot our results for the second derivatives 
of the scaling function with respect to the scaling variables. Here the statistical
errors are quite large and grow rapidly with the thickness of the film.
Therefore  we give only results for $L_0=8.5$ and $12.5$.  In the whole range
$x<x_c$ we find that $\theta^{(1,1)} \approx \theta^{(2,0)}$. The functions are 
negative and the amplitude increases rapidly as $x_c$ is approached.

Our results can be compared  with those of \cite{VaMaDi11}, who studied films with finite 
values of $h_1$ and $h_2$. In particular  in their Figs.~7  and 8 they 
give results for  $h_1=|h_2|$ and $h_2=0$, respectively. Their results for small $\tilde h_1$ 
are essentially consistent with ours. Matching their data with ours we get
$\tilde h_1 \approx 0.9 x_{h_1}$ for the relation between the scaling variables
that are used.
\section{Conclusions and outlook}
\label{conclus}
We study the thermodynamic Casimir force by using Monte Carlo 
simulations of lattice models. In particular we are concerned with the bulk universality 
class of the three-dimensional Ising model, which for example characterizes a continuous demixing transition  
of fluid binary mixtures.
In ref. \cite{mysphere} we used the exchange cluster algorithm, or geometric cluster algorithm
\cite{HeBl98}, to study the thermodynamic Casimir force between a spherical object and a plane 
substrate. 
The main point of the 
exchange cluster algorithm applied to this problem is that it allows to define a variance 
reduced estimator for the difference of the internal energy of two systems that are characterized
by slightly different distances between the spherical object and the substrate. In the 
case of the sphere-plate geometry it turned out to be mandatory to use this variance 
reduced estimator to get a meaningful result for the  thermodynamic Casimir force by using 
the approach discussed by Hucht \cite{Hucht}.  

Here, we go one step back and apply  the exchange cluster algorithm
to the film or plate-plate geometry. For this geometry, quite satisfactory numerical results
were obtained already. A long list of references is given in the introduction. 
We simulate the improved Blume-Capel model on the simple cubic lattice with 
$(+,+)$,  $(+,-)$, $(O,+)$, and $(O,O)$ boundary conditions, where $+$ and $-$ are
strongly symmetry breaking boundary conditions  and $O$ stands for the ordinary surface
universality class. For a discussion of these boundary conditions see the introduction and 
section \ref{themodel}. We demonstrate that also for the film geometry, the  exchange cluster
algorithm allows for a considerable reduction of the variance. The only exception are 
films with $(O,O)$ boundary conditions in the direct neighborhood of the transition of the
film.  This allowed us to simulate films with a larger thickness than before, allowing us
to consolidate previous results.
Our final estimates for the thermodynamic Casimir force only moderately improve on previous 
estimates. This is due to the fact that the remaining errors mainly stem from quantities
like $L_s$, see section \ref{themodel},  and $l_{ex,nor,0}$, see eq.~(\ref{defxh}), 
that were used as input.  These quantities were taken from previous work and are computed
by analyzing physical quantities different from the thermodynamic Casimir force.

In section \ref{algorithm} we discuss that the exchange cluster algorithm can be applied to a larger
class of boundary conditions than simulated here. In particular enhanced surface couplings
or  finite surface fields could be studied.   
Quite recently the authors of \cite{VaDi13,CaJaHo14} computed the thermodynamic Casimir
force in the presence of an external bulk field. To this end, one can compute
the difference in the excess free energy per area by integrating the difference in the excess
magnetisation per area over the external field \cite{CaJaHo14}, where the difference is
taken for films of thickness $L_0+1/2$ and $L_0-1/2$. The integration is started
at a strong external field, where the difference in the excess free energy vanishes.
Alternatively, one might start at a vanishing  external field, where the 
difference in the excess free energy per area is known from previous simulations.
It seems likely that the exchange cluster algorithm allows to reduce the variance 
of the difference in the excess magnetisation in such studies.
Furthermore one could think of applications different
from the thermodynamic Casimir force. For example one could compute the free energy of
defects. It would be interesting to check whether the simulation of spin glass models
could be speeded up by exchanging spins between replica.

The emphasis of our physics analysis is on $(O,O)$ boundary conditions. Films with such 
boundary conditions are expected to undergo a second order phase transition in the 
universality class of the two-dimensional Ising model. This transition has been studied 
for the Ising model on the simple cubic lattice for thicknesses up to $L_0=14$ in 
ref. \cite{KiOhIt96}. Here we obtain accurate results for thicknesses up to 
$L_0=64$ using the finite size scaling approach discussed in ref. \cite{CaHa96}.
Our numerical results nicely confirm the expectation that the transition belongs to
the universality class of the two-dimensional Ising model. We compute the finite size  
scaling function $g_n$, eq.~(\ref{defgn}), that governs the finite size scaling behavior 
of the free energy density in the universality class of the two-dimensional Ising model
for $n=2$ and $\infty$. We show that finite $L$-effects in the thermodynamic Casimir force, 
where $L$ is the extension of the film in the transversal directions, are described by
$g_n$. In particular using $g_{\infty}$, our knowledge of the inverse transition 
temperature of the film and the numerical matching of the scaling variable, we extrapolate
our results for the thermodynamic Casimir force to $L \rightarrow \infty$. For details see section
\ref{casimiroo}.  This approach could also be applied to other types of boundary 
conditions that do not break the $\mathbb{Z}_2$-symmetry of the problem, in particular to periodic
boundary conditions or enhanced surface couplings that allow to study the special 
surface universality class.

\section{Acknowledgement}
This work was supported by the DFG under the grant No HA 3150/3-1.

\end{document}